\newcommand\submitms{y}		

\documentclass[apj]{emulateapj}
\usepackage{apjfonts}

\usepackage{ifthen}
\usepackage{natbib}
\usepackage{amssymb, amsmath}
\usepackage{appendix}
\usepackage{etoolbox}

\bibpunct[, ]{(}{)}{,}{a}{}{,}

\usepackage[
bookmarks=true,           
bookmarksnumbered=true,   
colorlinks=true,          
citecolor=blue,           
linkcolor=blue,           
menucolor=blue,           
urlcolor=blue,            
linkbordercolor={0 0 1},  
pdfborder={0 0 1},
frenchlinks=true]{hyperref}

\providecommand{\adsurl}[1]{\href{#1}{ADS}}

\makeatletter
\patchcmd{\NAT@citex}
  {\@citea\NAT@hyper@{%
     \NAT@nmfmt{\NAT@nm}%
     \hyper@natlinkbreak{\NAT@aysep\NAT@spacechar}{\@citeb\@extra@b@citeb}%
     \NAT@date}}
  {\@citea\NAT@nmfmt{\NAT@nm}%
   \NAT@aysep\NAT@spacechar\NAT@hyper@{\NAT@date}}{}{}

\patchcmd{\NAT@citex}
  {\@citea\NAT@hyper@{%
     \NAT@nmfmt{\NAT@nm}%
     \hyper@natlinkbreak{\NAT@spacechar\NAT@@open\if*#1*\else#1\NAT@spacechar\fi}%
       {\@citeb\@extra@b@citeb}%
     \NAT@date}}
  {\@citea\NAT@nmfmt{\NAT@nm}%
   \NAT@spacechar\NAT@@open\if*#1*\else#1\NAT@spacechar\fi\NAT@hyper@{\NAT@date}}
  {}{}
\makeatother

\makeatletter
\DeclareRobustCommand{\lowcase}[1]{\@lowcase#1\@nil}
\def\@lowcase#1\@nil{\if\relax#1\relax\else\MakeLowercase{#1}\fi}
\makeatother

\newcommand\chisq{\ifmmode{\chi\sp{2}}\else\math{\chi\sp{2}}\fi}
\newcommand\redchisq{\ifmmode{ \chi\sp{2}\sb{\rm red}}
                    \else\math{\chi\sp{2}\sb{\rm red}}\fi}

\DeclareSymbolFont{UPM}{U}{eur}{m}{n}
\DeclareMathSymbol{\umu}{0}{UPM}{"16}
\let\oldumu=\umu
\renewcommand\umu{\ifmmode\oldumu\else\math{\oldumu}\fi}

\let\oldsim=\sim
\renewcommand\sim{\ifmmode\oldsim\else\math{\oldsim}\fi}
\let\oldpm=\pm
\renewcommand\pm{\ifmmode\oldpm\else\math{\oldpm}\fi}
\newcommand\by{\ifmmode\times\else\math{\times}\fi}

\newbox{\wdbox}
\renewcommand\c{\setbox\wdbox=\hbox{,}\hspace{\wd\wdbox}}
\renewcommand\i{\setbox\wdbox=\hbox{i}\hspace{\wd\wdbox}}

\newcount\timect
\newcount\hourct
\newcount\minct
\newcommand\now{\timect=\time \divide\timect by 60
         \hourct=\timect \multiply\hourct by 60
         \minct=\time \advance\minct by -\hourct
         \number\timect:\ifnum \minct < 10 0\fi\number\minct}

\catcode`@=11

\newcommand\comment[1]{}

\newcommand\commenton{\catcode`\%=14}
\newcommand\commentoff{\catcode`\%=12}

\renewcommand\math[1]{$#1$}
\newcommand\mathshifton{\catcode`\$=3}
\newcommand\mathshiftoff{\catcode`\$=12}

\comment{alignment tab}
\let\atab=&
\newcommand\atabon{\catcode`\&=4}
\newcommand\ataboff{\catcode`\&=12}

\let\oldmsp=\sp
\let\oldmsb=\sb
\def\sp#1{\ifmmode
           \oldmsp{#1}%
         \else\strut\raise.85ex\hbox{\scriptsize #1}\fi}
\def\sb#1{\ifmmode
           \oldmsb{#1}%
         \else\strut\raise-.54ex\hbox{\scriptsize #1}\fi}
\newbox\@sp
\newbox\@sb
\def\sbp#1#2{\ifmmode%
           \oldmsb{#1}\oldmsp{#2}%
         \else
           \setbox\@sb=\hbox{\sb{#1}}%
           \setbox\@sp=\hbox{\sp{#2}}%
           \rlap{\copy\@sb}\copy\@sp
           \ifdim \wd\@sb >\wd\@sp
             \hskip -\wd\@sp \hskip \wd\@sb
           \fi
        \fi}
\def\msp#1{\ifmmode
           \oldmsp{#1}
         \else \math{\oldmsp{#1}}\fi}
\def\msb#1{\ifmmode
           \oldmsb{#1}
         \else \math{\oldmsb{#1}}\fi}
\def\supon{\catcode`\^=7}
\def\supoff{\catcode`\^=12}
\def\subon{\catcode`\_=8}
\def\suboff{\catcode`\_=12}
\def\supsubon{\supon \subon}
\def\supsuboff{\supoff \suboff}

\newcommand\actcharon{\catcode`\~=13}
\newcommand\actcharoff{\catcode`\~=12}

\newcommand\paramon{\catcode`\#=6}
\newcommand\paramoff{\catcode`\#=12}

\comment{And now to turn us totally on and off...}

\newcommand\reservedcharson{ \commenton  \mathshifton  \atabon  \supsubon 
                             \actcharon  \paramon}

\newcommand\reservedcharsoff{\commentoff \mathshiftoff \ataboff \supsuboff 
                             \actcharoff \paramoff}

\catcode`@=12
\reservedcharsoff

\reservedcharson

\comment{ Must have ONLY ONE of these... trust these macros, they work

}
\reservedcharsoff

\actcharon
\if\submitms y

\else

\comment{
\received{}
\revised{}
\accepted{}
}
\fi

\reservedcharson

\shorttitle{TEA Code}
\shortauthors{Blecic {\em et al.}}

\begin{document}

\slugcomment{Submitted to {\em ApJ Supplement Series}.}

\title {TEA: A code for calculating thermochemical equilibrium abundances}

\author{Jasmina Blecic\altaffilmark{1}, Joseph
  Harrington\altaffilmark{1}, M. Oliver Bowman \altaffilmark{1}}

\affil{\sp1 Planetary Sciences Group, Department of Physics,
  University of Central Florida, Orlando, FL 32816-2385, USA}

\email{jasmina@physics.ucf.edu}

\begin{abstract}

We present an open-source Thermochemical Equilibrium Abundances (TEA)
code that calculates the abundances of gaseous molecular species.  The
code is based on the methodology of
\citet{WhiteJohnsonDantzig1958JGibbs} and \citet{Eriksson1971}. It
applies Gibbs free-energy minimization using an iterative, Lagrangian
optimization scheme. Given elemental abundances, TEA calculates
molecular abundances for a particular temperature and pressure or a
list of temperature-pressure pairs. We tested the code against the
method of \citet{BurrowsSharp1999apjchemeq}, the free thermochemical
equilibrium code CEA (Chemical Equilibrium with Applications), and the
example given by \citet{WhiteJohnsonDantzig1958JGibbs}. Using their
thermodynamic data, TEA reproduces their final abundances, but with
higher precision. We also applied the TEA abundance calculations to
models of several hot-Jupiter exoplanets, producing expected
results. TEA is written in Python in a modular format. There is a 
start guide, a user manual, and a code document in addition to this
theory paper. TEA is available under a reproducible-research, 
open-source license via {\tt https://github.com/dzesmin/TEA}.


\end{abstract}

\section{INTRODUCTION}
\label{intro}

There are two methods to calculate chemical equilibrium: using
equilibrium constants and reaction rates, i.e., kinetics, or
minimizing the free energy of a system \citep{bahn1960kinetics,
  ZeleznikGordon:1968}.

The kinetic approach, where the pathway to equilibrium needs to be
determined, is applicable for a wide range of temperatures and
pressures. However, using kinetics for thermochemical equilibrium
calculations can be challenging. Chemical equilibrium can be
calculated almost trivially for several reactions present in the
system, but as the number of reactions increases, the set of numerous
equilibrium constant relations becomes hard to solve
simultaneously. To have an accurate kinetic assessment of the system,
one must collect a large number of reactions and associate them with
the corresponding rates. This is not an issue at lower temperatures,
where reaction rates are well known. However, at high temperatures,
where thermochemical equilibrium should prevail, one needs to know
forward and reverse reactions and corresponding reaction rates, which
are less well known.

The advantage of the free energy minimization method is that each
species present in the system can be treated independently without
specifying complicated sets of reactions a priori, and therefore, a
limited set of equations needs to be solved
\citep{ZeleznikGordon:1960}. In addition, the method requires only
knowledge of the free energies of the system, which are well known,
tabulated, and can be easily interpolated or extrapolated.

Thermochemical equilibrium calculations have been widely used in
chemical engineering to model combustion, shocks, detonations and the
behaviour of rockets and compressors \citep[e.g.,
][]{MillerEtal-annualReview, BelfordStrehlow-annualReview}. In
astrophysics, they have been used to model the solar nebula, the
atmospheres and circumstellar envelopes of cool stars, and the
volcanic gases on Jupiter's satellite Io \citep[e.g.,
][]{LaurettaLoddersFegley997S-nebula, LoddersFegley1993-circum,
  ZolotovFegley1998-Io}. Thermochemistry also governs atmospheric
composition in vast variety of giant planets, brown dwarfs, and
low-mass dwarf stars \citep[][and references therein ]{Lodders02,
  Visscher2006, VisscherEtal2010-chem}.

\subsection{Chemical Models of Exoplanets}
\label{sec:GibbsMinim}

To perform a comprehensive study of a planetary atmosphere, aside from
thermoequilibrium chemistry, one must consider disequilibrium
processes like photochemistry, vertical mixing, horizontal transport,
and transport-induced quenching \citep{MosesEtal2011-diseq,
  MosesEtal2013-COratio, VenotEtal2012-chem, Venot2014-metallicity,
  Agundez2012, Agundez2014, LineEtal2010ApJHD189733b,
  LineEtal2011-kinetics, VisscherMoses2011-quench}. Today, we have 1D
chemical models that integrate thermochemistry, kinetics, vertical
mixing, and photochemistry \citep{LineEtal2011-kinetics,
  MosesEtal2011-diseq, VisscherEtal2010IcarJupiter}. These models have
an ability to smoothly transition from the thermochemical-equilibrium
regime to transport-quenched and photochemical regimes. Specifically,
in giant planets, we can distinguish three chemical layers: deep
within the planetary atmosphere, the temperatures and pressures are so
high that chemical reaction timescales are short, ensuring a chemical
equilibrium composition; at lower temperatures and pressures higher in
the atmosphere, the timescales for chemical reactions slows down,
reaching the vertical transport timescale and smoothing the vertical
mixing-ratio profile by producing quenched abundances; high in the
atmosphere, the host star's ultraviolet radiation destroys stable
molecules, driving photochemical reactions.
  
Photochemical models today face several difficulties. They lack
high-temperature photochemical data, and the list of reactions and
associated rate coefficients are not well defined or are conflicted
\citep{VenotEtal2012-chem, VisscherEtal2010IcarJupiter}. In addition,
the exoplanet photospheres observed with current instruments are
sampled within the region of the atmosphere dominated by vertical
mixing and quenching, but not by photochemistry
\citep{LineYung2013-diseq}.

The majority of early hot-Jupiter atmospheric models assumed chemical
composition consistent with thermochemical equilibrium \citep[e.g.,
][]{BurrowsEtal2007ApJHD209458, FortneyEtal2005apjlhjmodels,
  MarleyEtal2007ApJ, FortneyEtal2010ApJ-transmission,
  BurrowsSharp1999apjchemeq, SharpBurrows2007Apjopacities,
  RogersEtal2009ApJ-groundCoRoT1b}. More recently, a variety of 1D
photochemical models has been used to explore the compositions of hot
Jupiters \citep{MosesEtal2011-diseq, Zahnle09-SulfurPhotoch,
  LineEtal2010ApJHD189733b, KopparapuEtal2012-photoch,
  VenotEtal2012-chem, LineYung2013-diseq, Visscher2006,
  VisscherEtal2010IcarJupiter, VenotEtal2012-chem}. A common
conclusion of these studies is that in hot atmospheres (\math{T}
\math{>} 1200 K), disequilibrium effects are so reduced that
thermochemical equilibrium prevails.

Using secondary eclipse observations as the most fruitful technique
today to assess atmospheric composition \citep[e.g.,
][]{KnutsonEtal2009ApJ-Tres4Inversion,
  KnutsonEtal2009ApJ-redistribution, Machalek2008-XO-1b,
  StevensonEtal2012apjHD149026b, FraineEtal2013-GJ1214b,
  CrossfieldEatl2012ApJWASP12b-reavaluation,
  TodorovEtal2012ApJ-XO4b-HATP6b-HATP8b, DesertEtal2009ApJ-CO,
  DemingEtal2010arXiv-CoRoT12, Demory2007aaGJ436bspitzer,
  MadhusudhanSeager2009ApJ-AbundanceMethod,
  MadhusudhanEtal2011natWASP12batm, BlecicEtal2013-WASP14b,
  BlecicEtal2014-WASP43b}, \citet{LineYung2013-diseq} studied the most
spectroscopically active species in the infrared on eight hot planets
(GJ436b, WASP-12b, WASP-19b, WASP-43b, TrES-2b, TrES-3b, HD 189733b,
and HD 149026b), with equilibrium temperatures ranging between 744 K
and 2418 K. They chose to evaluate the presence of disequilibrium
chemistry at 100 mbar, where most secondary-eclipse observations
sample, i.e., where their thermal emission weighting functions usually
peak. They find that all of the models are consistent with
thermochemical equilibrium within 3\math{\sigma} (the work of
\citealp{StevensonEtal2010Natur}, however, questions this conclusion for
GJ436b). They also show that for the hottest planets, (T\sb{100mb}
\math{>} 1200 K ), CH\sb{4}, CO, H\sb{2}O, and H\sb{2} should be in
thermochemical equilibrium even under a wide range of vertical mixing
strengths.

Thermochemical equilibrium calculations are the starting point for
initializing models of any planetary atmosphere. In general,
thermochemical equilibrium governs the composition of the deep
atmospheres of giant planets and brown dwarfs, however, in cooler
atmospheres thermoequilibrium calculations are the necessary baseline
for further disequilibrium assessment. They can also provide a
first-order approximation for species abundances as a function of
pressure, temperature, and metallicity for a variety of atmospheres
\citep[e.g., ][]{VisscherEtal2010IcarJupiter, Lodders02}.

The Gibbs free energy minimization method for calculating
thermochemical equilibrium abundances of complex mixtures was first
introduced by \citet{WhiteJohnsonDantzig1958JGibbs}.  Prior to 1958
all equilibrium calculations were done using equilibrium constants of
the governing reactions. \citet{WhiteJohnsonDantzig1958JGibbs} were
the first to develop a method that makes no distinction among the
constituent species and does not need a list of all possible chemical
reactions and their rates. Rather, it depends only on the chemical
potentials of the species involved. To derive the numerical solution,
they apply two computational techniques: a steepest-descent method
applied to a quadratic fit and the linear programming method.

Following their methodology, \citet{Eriksson1971} developed the SOLGAS
code that calculates equilibrium composition in systems containing
ideal gaseous species and pure condensed phases. Subsequent
modification of this code were made by
\citet{eriksson1973thermodynamic}, \citet{eriksson1975thermodynamic},
and \citet{besmann1977solgasmix}, after which the code was modified
for astrophysical applications and called SOLAGASMIX by
\citet{SharpHuebner90, PetaevWood1998MPSConden},
\citet{BurrowsSharp1999apjchemeq}, and
\citet{SharpBurrows2007Apjopacities}.

The Gibbs free energy minimization approach has been used by many
authors in the exoplanetary field \citep[e.g., ][]{SeagerPhD1999,
  Seager2010-ExoplanetAtmospheres,
  MadhusudhanSeager2009ApJ-AbundanceMethod, MadhusudhanSeager2010,
  SharpHuebner90}. In addition to SOLAGASMIX and other proprietary
codes \citep[e.g., CONDOR by][]{FegleyLodders1994IcarChemiModelSatJup,
  LoddersFegley1994-CONDOR}, and one analytic method to calculate
major gaseous species in planetary atmospheres by
\citet{BurrowsSharp1999apjchemeq}, just one free-software code CEA,
\citep[Chemical Equilibrium with Applications, {\tt
    http://www.grc.nasa.gov/WWW/CEAWeb}, by][]{GordonMcBride:1994}, is
available to the exoplanet community.

In this paper, we present a new open-source code, Thermochemical
Equilibrium Abundances (TEA). The TEA code is a part of the
open-source Bayesian Atmospheric Radiative Transfer project ({\tt
  https://github.com/joeharr4/BART}). This project consists of three
major parts: TEA - this code, a radiative-transfer code that 
models planetary spectra, and a statistical module that compares
theoretical models with observations.

TEA calculates the equilibrium abundances of gaseous molecular
species. Given a single \math{T, P} point or a list of \math{T, P}
pairs (the thermal profile of an atmosphere) and elemental abundances,
TEA calculates mole fractions of the desired molecular species. The
code is based on the Gibbs free energy minimization calculation of
\citet{WhiteJohnsonDantzig1958JGibbs} and \citet{Eriksson1971}. TEA
uses 84 elemental species and the thermodynamical data for more then
600 gaseous molecular species available in the provided JANAF (Joint
Army Navy Air Force) tables \citep[{\tt
    http://kinetics.nist.gov/janaf/},][]{ChaseEtal1982JPhJANAFtables,
  ChaseEtal1986bookJANAFtables}. TEA can adopt any initial elemental
abundances. For user convenience a table with solar photospheric
elemental abundances from \citet{AsplundEtal2009-SunAbundances} is
provided.

The code is written in Python in an architecturally modular format. It
is accompanied by detailed documentation, a start guide, the TEA User 
Manual (Bowman and Blecic), the TEA Code Description document 
(Blecic and Bowman), and the TEA Theory document (this paper), so the
user can easily modify it. The code is actively maintained and available
to the scientific community via the open-source development website {\tt
  GitHub.com} ({\tt https://github.com/dzesmin/TEA,
  https://github.com/dzesmin/TEA-Examples}). This paper covers an
initial work on thermochemical calculations of species in gaseous
phases. Implementation of condensates is left for future work.

In this paper, we discuss the theoretical basis for the method applied
in the code. Section \ref{sec:GibbsMinim} explains the Gibbs Free
energy minimization method; Section \ref{sec:Lagrang} describes the
general Lagrangian optimization method and its application in TEA; in
Section \ref{sec:lambda} we introduce the Lambda Correction algorithm
for handling negative abundances that follow from the Lagrangian
method; Section \ref{sec:struct} describes the layout of the TEA code;
Section \ref{sec:applic} explores chemical equilibrium abundance
profiles of several exoplanetary atmospheres; Section
\ref{sec:validity} compares our code to other methods available, and
Section \ref{sec:conc} states our conclusions.

\section{Gibbs Free Energy Minimization Method}
\label{sec:GibbsMinim}

Equilibrium abundances can be obtained by using different combinations
of thermodynamical state functions: temperature and pressure --
(\math{t, p}), enthalpy and pressure -- (\math{H, p}), entropy and
pressure -- (\math{S, p}), temperature and volume -- (\math{t, v}),
internal energy and volume -- (\math{U, v}), etc. Depending on how the
system is described, the condition for equilibrium can be stated in
terms of Gibbs free energy, helmholtz energy, or entropy. If a
thermodynamic state is defined with temperature and pressure, Gibbs
free energy (G) is most easily minimized, since those two states are
its natural, dependent variables.

Gibbs free energy represents a thermodynamic potential that measures
the useful work obtainable by the system at a constant temperature and
pressure. Thus, the Gibbs free energy minimization method minimizes
the total chemical potential of all involved species when the system
reaches equilibrium.

The Gibbs free energy of the system at a certain temperature is the
sum of the Gibbs free energies of its constituents:

\begin{equation}
G_{sys}(T) = \sum_{i}^nG_{i}(T) \, ,
\label{Gibbs-sys}
\end{equation}

\noindent where \math{G\sb{sys}(T)} is the total Gibbs free energy of
the system for \math{n} chemical species, \math{G\sb{i}(T)} is the
Gibbs free energy of a gas species \math{i}, and \math{T} is the
temperature. The total Gibbs free energy of the system is expressed as
the sum of the number of moles \math{x} of the species \math{i},
\math{x\sb{i}}, and their chemical potentials \math{g\sb{i}(T)}:


\begin{equation}
G_{sys}(T) = \sum_{i}^nx_{i}\,g_{i}(T) \, .
\label{Gibbs-pot}
\end{equation}

\noindent The chemical potential \math{g\sb{i}(T)} depends on the
chemical potential at the standard state \math{g\sp{0}\sb{i}(T)} and
the activity \math{a\sb{i}},

\begin{equation}
g_{i}(T) = g_{i}^0 (T)+ RT\ln a_{i} \, ,
\label{pot}
\end{equation}

\noindent where \math{R} is the gas constant, \math{R} =
\math{k\sb{B}N\sb{A}}, and \math{k\sb{B}} and \math{N\sb{A}} are the
Boltzmann constant and Avogadro's number, respectively. Activities
for gaseous species, which are treated as ideal, are equal to the
partial pressures, and for condensates they equal 1:

\begin{equation}
a_{i} = P_{i} = P\,\frac{x_{i}}{N} \, , \, \,\,\,\, {\rm for\,gases}
\label{pot1}
\end{equation}

\begin{equation}
a_{i} = 1 \,, \, \,\,\,\,\,\,\,\,{\rm for \,condensates} \, ,
\label{pot2}
\end{equation}

\noindent where \math{P} is the total pressure of the atmosphere, \math{N}
is the total number of moles of all species involved in the
system. Hence, Equation (\ref{pot}) for gaseous species becomes:

\begin{equation}
g_{i}(T) = g_{i}^0(T) + RT\ln P_i \, .
\label{pot-pot}
\end{equation}

\noindent Combining Equation (\ref{pot-pot}) with Equation
(\ref{Gibbs-pot}), the Gibbs free energy of the system becomes:

\begin{equation}
G_{sys}(T) = \sum_{i}^n x_{i}\Big(g_{i}^0(T) + RT\ln P_i\Big) \, ,\\
\label{Gibbsfree2}
\end{equation}

\noindent or,

\begin{equation}
G_{sys}(T) = \sum_{i}^n\,x_{i}\Big(g_{i}^0(T) + RT\ln P + RT
\ln\,\frac{x_{i}}{N}\Big) \, ,
\label{GibbsFull}
\end{equation}

\noindent For our purposes, it is more convenient to write Equation
(\ref{GibbsFull}) in unitless terms:

\begin{eqnarray}
\label{eq:eqmin}
\frac{G_{sys}(T)}{RT} = \sum_{i=1}^n x_{i} \Big[\frac{g_{i}^0(T)}{RT}
  + \ln P + \ln\frac{x_{i}}{N}\Big]\, .
\end{eqnarray}

Equation (\ref{eq:eqmin}) requires a knowledge of the free energy of
each species as a function of temperature. These can be obtained from
the JANAF tables \citep[{\tt http://kinetics.nist.gov/janaf/},
][]{ChaseEtal1982JPhJANAFtables, ChaseEtal1986bookJANAFtables,
  BurrowsSharp1999apjchemeq}, or easily derived from other tabulated
functions.

To extract free energies, \math{g\sb{i}\sp{0}(T)/RT}, from the JANAF
tables, we used the expression given in \citet[][]{Eriksson1971},
Equation (2):

\begin{eqnarray}
\label{eq:JANAFconv}
\frac{g_{i}^0(T)}{RT} = 1/R\Big[\frac{G_{i}^0 - H_{298}^0}{T}\Big] +
\frac{\Delta_f H_{298}^0}{RT}\, ,
\end{eqnarray}

\noindent where \math{g\sb{i}\sp{0}(T)} is given in J/mol, \math{R} =
8.3144621 J/K/mol, \math{H\sb{298}\sp{0}} is the enthalpy (heat
content) in the thermodynamical standard state at a reference
temperature of 25\sp{o}C = 298.15 K, \math{G\sb{i}\sp{0}} is the Gibbs
free energy in J/mol, (\math{G\sb{i}^0 - H\sb{298}\sp{0}/T}) is the
free-energy function in J/K/mol, and \math{\Delta\sb{f}
  H\sb{298}\sp{0}} is the heat of formation at 298.15 K in
kJ/mol. Thus, our conversion equation becomes:

\begin{eqnarray}
\label{eq:JANAFconvFinal}
\frac{g_{i}^0(T)}{RT} = 1/R\Big[\frac{G_{i}^0 - H_{298}^0}{T}\Big] +
\frac{\Delta_f H_{298}^0 1000}{RT}\, ,
\end{eqnarray}

\noindent \math{G\sb{i}^0 - H\sb{298}\sp{0}/T} is the fourth term in
the JANAF tables and \math{\Delta\sb{f} H\sb{298}\sp{0}} is the
sixth. The free energy function of a species corresponding to a
temperature other than those provided in the JANAF tables is
calculated using spline interpolation.

Alternatively, the free energies can be calculated using the eighth
term in the JANAF tables, following Equation (3) from
\citet[][]{Eriksson1971}:

\begin{eqnarray}
\label{eq:JANAFconvFinal2}
\frac{g_{i}^0(T)}{RT} = -\,ln\,(10)\,log_{10}\,(K_{f})\, ,
\end{eqnarray}

\noindent where \math{K\sb{f}} is the equilibrium constant of
formation.

To determine the equilibrium composition, we need to find a
non-negative set of values \math{x\sb{i}} that minimizes Equation
(\ref{eq:eqmin}) and satisfies the mass balance constraint:

\begin{eqnarray}
\sum_{i=1}^n a_{ij}\, x_{i} = b_{j}\,, \,\,\,\,(j = 1, 2, ...,m)\, ,
\label{masbal}
\end{eqnarray}

\noindent where the stoichiometric coefficient \math{a\sb{ij}}
indicates the number of atoms of element \math{j} in species \math{i}
(e.g., for CH\sb{4} the stoichiometric coefficient of C is 1 and the
stoichiometric coefficient of H is 4), and \math{b\sb{j}} is the total
number of moles of element \math{j} originally present in the mixture.

We use the reference table containing elemental solar abundances given
in \citet{AsplundEtal2009-SunAbundances} Table 1 for \math{b}
values. \citet{AsplundEtal2009-SunAbundances} adopt the customary
astronomical scale for logarithmic abundances, where hydrogen is
defined as log \math{\epsilon{H}} = 12.00, and log \math{\epsilon{X}}
= log(\math{N\sb{X}/N\sb{H}})+12, where \math{N\sb{X}} and
\math{N\sb{H}} are the number densities of element \math{X} and
\math{H}, respectively. Thus, their values are given in {\em dex}
(decimal exponent) units. We transform these values into elemental
fractions by number, i.e., ratio of number densities. We convert each
species dex elemental abundance into number density and divide it by
the hydrogen number density \citep[Section
  3]{AsplundEtal2009-SunAbundances}. The final output are fractional
abundances (mole mixing fractions), i.e., the ratio of each species'
number of moles to the number of moles in the mixture.

\section{Lagrangian Method of Steepest Descent}
\label{sec:Lagrang}

To find equilibrium abundances of the desired molecular species at a
given temperature and pressure, we need to minimize Equation
(\ref{eq:eqmin}). To do so, we have to apply a technique that
minimizes a multi-variate function under constraint. There are many
optimization techniques used to find the minima of a function subject
to equality constraints (e.g., line search method, Dantzig-simplex
method for linear programming, Newton-Raphson method,
Hessian-conjugate gradient method, Lagrangian steepest-descent
method). The main advantage of the Lagrangian steepest-descent method
is that the number of equations to solve scales with the number of
different types of atoms present in the mixture, which is usually a
much smaller number than the possible number of molecular
constituents. This allows the code to be executed much faster than in
other methods.

Gradient descent, also known as steepest descent, is an algorithm for
finding a local minimum of a function. At each iteration, the method
takes steps towards the minimum, where each step is proportional to
the negative gradient of the function at the current point. If a
function \math{f(x)} is defined and differentiable in the neighborhood
of a point \math{a}, then \math{f(x)} decreases most rapidly in the
direction of the negative gradient, \math{ - \nabla f(\math{a})}. From
this, it follows that if \math{b} = \math{a - \lambda\nabla
  f(\mathbf{a})}, then \math{f(a) > f(b)} if \math{\lambda} is small
enough. Starting with a guess \math{x\sb{0}} for a local minimum of
\math{f}, and considering a sequence \math{x\sb{0}, x\sb{1},
  x\sb{2},...} such that \math{x\sb{n+1} = x\sb{n} - \lambda\nabla
  f(x\sb{n}),\, n \geq 0}, one gets \math{f(x\sb{0})} \math{\geq}
\math{f(x\sb{1})} \math{\geq} \math{f(x\sb{2})} \math{\geq}... . This
sequence of \math{x\sb{n}} converges to a desired local minimum if the
correct \math{\lambda} value is assigned. The value of \math{\lambda}
can vary at each iteration. If the function \math{f} is convex, the
local minimum is also the global minimum.

\begin{figure*}[ht]
\centerline{
\includegraphics[height=5cm, clip]{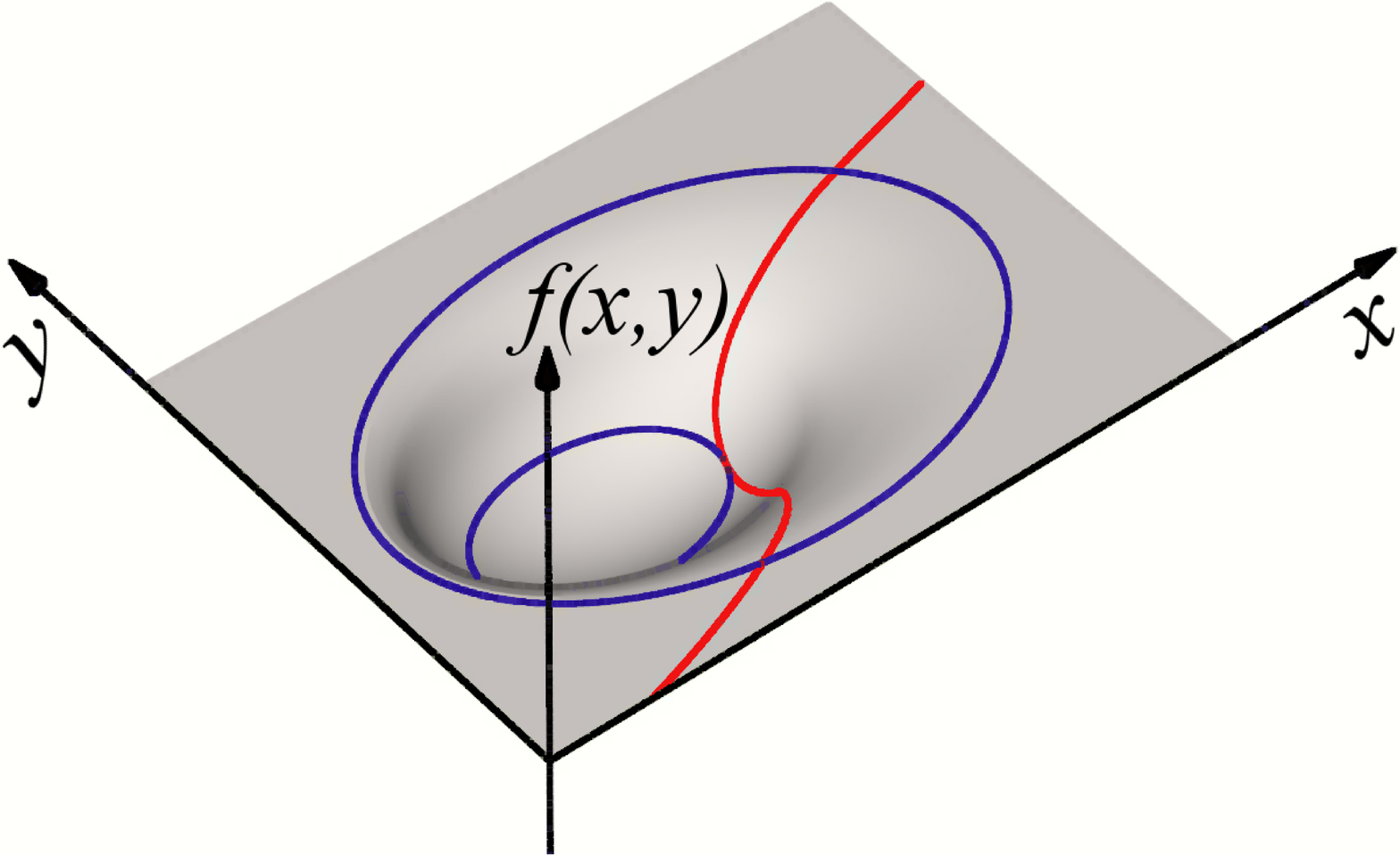}\hspace{35pt}
\includegraphics[height=5.1cm, clip]{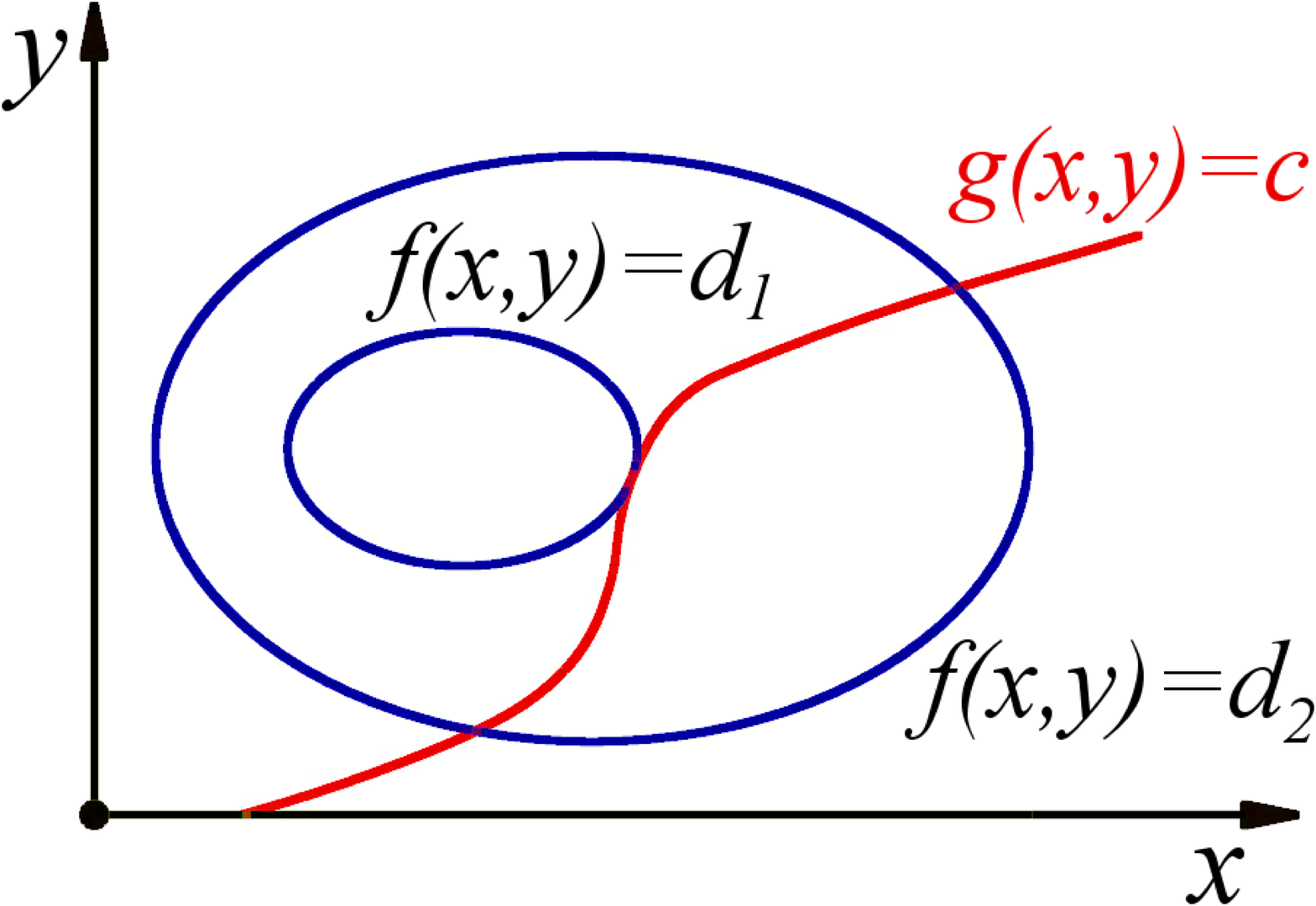}}
\caption{ Example of the Lagrangian minimization approach.  {\bf
    Left:} The 3D illustration of the minimization problem. Blue lines
  indicate starting and ending values of \math{f(x,y)} during
  minimization. An (\math{x}, \math{y}) pair is found that minimizes
  \math{f(x,y)} (bottom blue line) subject to a constraint
  \math{g(x,y)=C} (red line).  {\bf Right:} Contour map of the left
  figure. The point where the red line (constraint) tangentially
  touches a blue contour is the solution. Since \math{d_1<d_2}, the
  solution is the minimum of \math{f(x,y)}.}
\label{fig:Corr-Hist}
\end{figure*}

Our code implements a more complex version of the method outlined
above. The problem consists of some function \math{f(x,y)} subject to
a constraint \math{g(x, y)=C}. In this case, we need both \math{f} and
\math{g} to have continuous first partial derivatives. Thus, we
introduce a new variable called the {\em Lagrangian multiplier},
\math{\pi}, where:

\begin{eqnarray}
\Lambda (x, y, \lambda) = f(x, y) \pm \pi\, (g\,(x, y) - C)\, ,
\label{Lagr-mult}
\end{eqnarray}

\noindent which allows us to find where the contour of \math{g(x,
  y)=C} tangentially touches \math{f(x,y)} (Figure
\ref{fig:Corr-Hist}). The point of contact is where their gradients
are parallel:

\begin{eqnarray}
\nabla_{xy}\,f(x, y) = - \pi\nabla_{xy}\,g(x, y) \, .
\label{Lagr-cond}
\end{eqnarray}

\noindent The constant \math{\pi} allows these gradients to have
different magnitudes. To find the minimum, we need to calculate all
partial derivatives of the function \math{\Lambda}, equate them with
zero,

\begin{eqnarray}
\nabla_{x, y, \pi}\,\Lambda(x, y, \pi) = 0 \, ,
\label{Lagr-final-cond}
\end{eqnarray}

\noindent and follow the same iteration procedure as explained above.

\subsection{Lagrangian Method in TEA}
\label{sec:LagrTEA}

To implement this in our code, we followed the methodology derived in
\citet{WhiteJohnsonDantzig1958JGibbs}. We applied an iterative
solution to the energy minimization problem, where the mole numbers of
the desired molecular species are recomputed at each step and the new
direction of steepest descent is calculated. This produces improved
mole number values, which however, could be negative. Thus, two short
procedures are required in each iteration cycle: solving a set of
simultaneous linear equations for an improved direction of descent
(described in this Section) and approximately minimizing a convex
function of one variable, \math{\lambda}, to ensure that all improved
mole number values are positive (Section \ref{sec:lambda}).

To calculate the direction of steepest descent (following the
methodology derived in Section \ref{sec:Lagrang}) and initiate the
first iteration cycle, we first need to solve the mass balance
equation, Equation (\ref{masbal}). We start from any positive set of
values for the initial mole numbers, \math{y = (y\sb{1}, y\sb{2}, ...,
  y\sb{n})}, as our initial guess:

\begin{eqnarray}
\sum_{i=1}^n a_{ij}\, y_{i} = b_{j}\, \,\,\,\,(j = 1, 2, ...,m) \, .
\label{masbal2}
\end{eqnarray}

\noindent To satisfy the mass balance Equation (\ref{masbal2}), some
\math{y\sb{i}} variables must remain as free parameters. In solving
these equations, we leave as many free parameters as we have elements
in the system, thus ensuring that the mass balance equation can be
solved for any number of input elements and output species the user
chooses. We set all other \math{y\sb{i}} to a known, arbitrary
number. Initially, the starting values for the known species are set
to 0.1 moles, and the mass balance equation is calculated. If that
does not produce all positive mole numbers, the code automatically
sets known parameters to 10 times smaller and tries again. The initial
iteration input is set when all mole numbers are positive, and the
mass balance equation is satisfied.

To follow with the Lagrangian method, we denote two terms in Equation
 (\ref{eq:eqmin}) as:

\begin{eqnarray}
\label{eq:c}
c_i = \frac{g_{i}^0(T)}{RT} + \ln P \, ,
\end{eqnarray}

\noindent where \math{P} is the pressure in bar. Using \math{c\sb{i}},
we denote the right side of Equation (\ref{eq:eqmin}) as the variable
\math{f\sb{i}(Y)}:

\begin{eqnarray}
f_i(Y) = y_i\Big[c_i + \ln\frac{y_{i}}{\bar{y}}\Big]\, ,
\label{f}
\end{eqnarray}

\noindent where \math{Y} = (\math{y\sb{1}, y\sb{2}, ..., y\sb{n}}) and
\math{\bar{y}} is the total initial number of moles.  The left 
side of Equation (\ref{eq:eqmin}), \math{G\sb{sys}(T)/RT}, we denote
as function \math{F(Y)}:

 \begin{eqnarray}
F(Y) = \sum_{i=1}^n\,y_i\Big[c_i + \ln\frac{y_{i}}{\bar{y}}\Big]\, .
\label{sumf}
\end{eqnarray}

Then, we do a Taylor series expansion of the function \math{F} about
\math{Y}. This yields a quadratic approximation \math{Q(X)}:

 \begin{eqnarray}
Q(X) = F(X)\Big|_{X=Y} + \sum_i \frac{\partial F}{\partial
  x_i}\Big|_{X=Y} \, \Delta_i + \\ \nonumber
\frac{1}{2}\sum_i\sum_k\frac{\partial^2\,F}{\partial x_i\partial
  x_k}\Big|_{X=Y}\, \Delta_i\, \Delta_k\, .
\label{quadrApp}
\end{eqnarray}

\noindent where \math{\Delta\sb{i} = x\sb{i} - y\sb{i}}, and
\math{x\sb{i}} are the improved mole numbers. This function is
minimized using the Lagrangian principle. We now introduce Lagrangian
multipliers as \math{\pi\sb{j}}:

 \begin{eqnarray}
G(X) = Q(X) + \sum_j \pi_j (- \sum_i a_{ij}\,x_i +b_j)\, ,
\label{minim}
\end{eqnarray}

\noindent and calculate the first derivatives, \math {\partial G/
  \partial x\sb{i}}, of the new function. We equate them to zero to
find the minima, \math {\partial G/ \partial x\sb{i}} = 0.

We solve for \math{x\sb{i}} from Equation (\ref{minim}) by combining
Equation (\ref{masbal2}) and (\ref{f}) with the fact that
\math{\bar{x}} is the sum of the improved mole numbers, \math{\bar{x}
  = \sum_{i=1}^n x_i}. The improved number of moles, \math{x\sb{i}},
are given as:

\begin{eqnarray}
x_i = - f_i(Y) + (\frac{y_{i}}{\bar{y}})\,\bar{x} + (\sum_{j=1}^m
\pi_j\,a_{ij})\,y_i\, ,
\label{x_i2}
\end{eqnarray}

\noindent while the Lagrangian multipliers, \math{\pi\sb{j}}, are
expressed as:

\begin{eqnarray}
\sum_{j=1}^m \pi_j\,\sum_{i=1}^n\,a_{ij}\,y_i = \sum_{i=1}^n y_i
\Big[\frac{g_{i}^0(T)}{RT} + \ln P +\ln {\frac{y_{i}}{\bar{y}}}\Big]
\, ,
\label{pi}
\end{eqnarray}

\noindent where \math{j} iterates over the \math{m} elements and
\math{i} iterates over the \math{n} species. \math{\bar{x}} and
\math{\bar{y}} are the sums of improved and initial number of
moles, respectively. Using Equation (\ref{f}), we can now rewrite Equation
(\ref{pi}) as:

\begin{eqnarray}
\sum_{j=1}^m \pi_j\,b_j = \sum_{i=1}^n f_i(Y) \, .
\label{pi2}
\end{eqnarray}

If we further denote the constants with:

\begin{eqnarray}
r_{jk} = r_{kj} = \sum_{i=1}^n (a_{ij}\,a_{ik})\,y_{i}, 
\label{r}
\end{eqnarray}

\noindent combining Equations (\ref{x_i2}), (\ref{pi2}), and (\ref{r}),
we get the following system of \math{m+1} equations that can easily be
solved:


\begin{align}
  r_{11}\pi_1 + r_{12}\pi_2 + ... + r_{1m}\pi_m + b_1\,u =
  \sum_{i=1}^n a_{i1}\,f_i(Y)\, , \nonumber \\ r_{21}\pi_1 +
  r_{22}\pi_2 + ... + r_{2m}\pi_m + b_2\,u = \sum_{i=1}^n
  a_{i2}\,f_i(Y)\, , \nonumber \\ . \hspace{120pt} , \nonumber
  \\ . \hspace{120pt} , \label{final-set} \\ . \hspace{120pt} ,
  \nonumber \\ r_{m1}\pi_1 + r_{m2}\pi_2 + ... + r_{mm}\pi_m + b_m\,u
  = \sum_{i=1}^n a_{im}\,f_i(Y)\, , \nonumber \\ b_1\pi_1 + b_2\pi_2 +
  ... + b_m\pi_m + 0\,u = \sum_{i=1}^n f_i(Y)\, , \nonumber
\end{align}

\noindent where:

\begin{eqnarray}
u = -1 + \bar{x}/\bar{y}\, .
\label{u}
\end{eqnarray}

The solutions to Equations (\ref{final-set}) and (\ref{u}) will give
\math{\pi\sb{j}} and \math{u}, and from them using Equation
(\ref{x_i2}) we can calculate the next set of improved mole numbers,
i.e., an improved direction of descent, \math{\Delta\sb{i} = x\sb{i} -
  y\sb{i}}.

\section{Lambda Correction Algorithm}
\label{sec:lambda}

Solving a system of linear equations (i.e.,
performing the Lagrangian calculation) can also lead to negative mole
numbers for some species, so a short
additional step is needed to eliminate this possibility and guarantee
a valid result.

\begin{figure}[!h]
\centering
\includegraphics[height=4.5cm, clip]{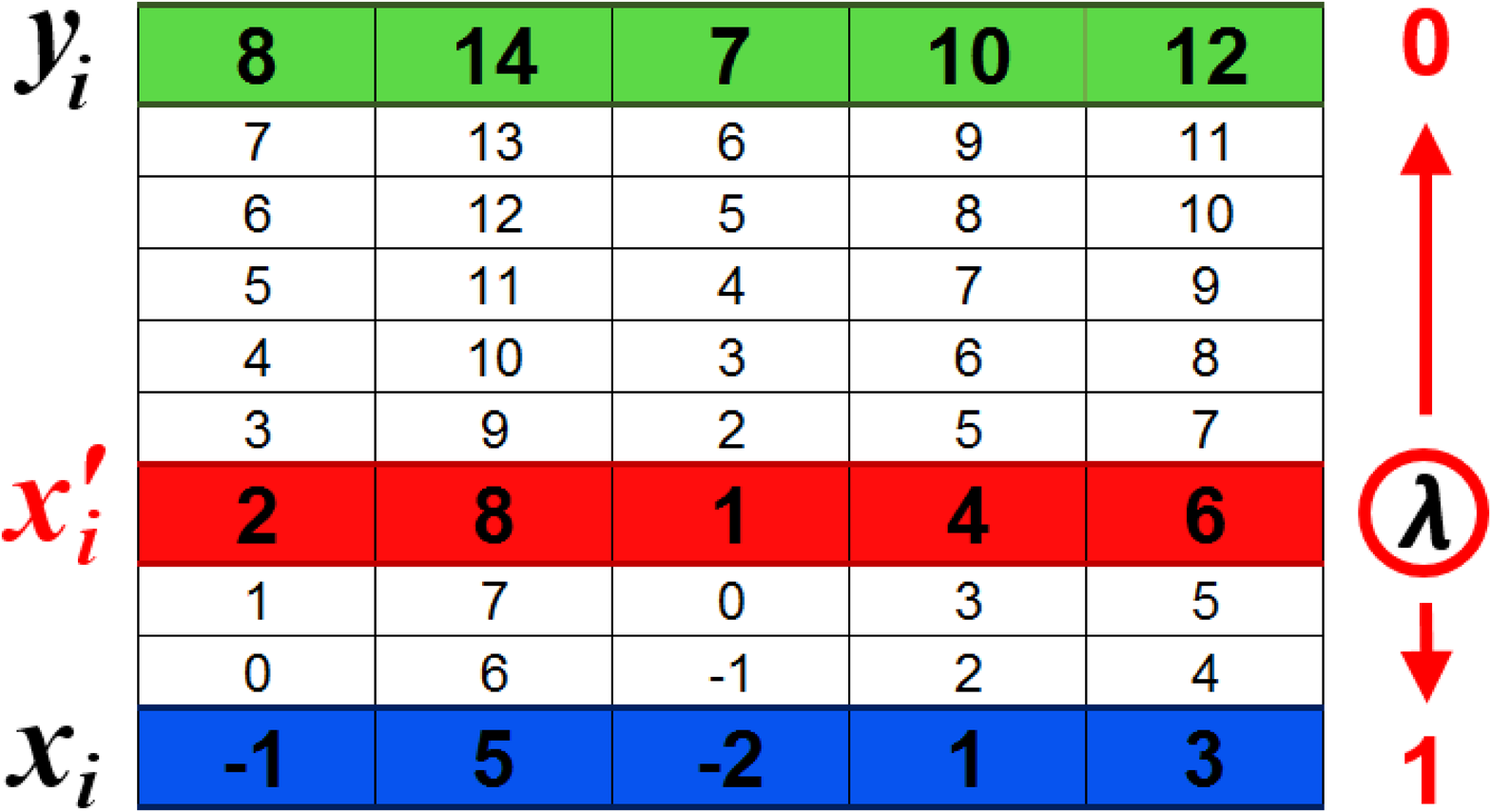}
\caption{Simplified illustration of the lambda correction algorithm.
  Initial values for one hypothetical Lagrangian iteration cycle,
  \math{y\sb{i}}, are given in green. These values are all positive
  and satisfy the mass balance equation, Equation \ref{masbal2}. The
  {x\sb{i}} values, given in blue, are the values produced by the
  Lagrangian calculation. These values can be negative, but they also
  satisfy the mass balance equation.  The \math{x\sp{'}\sb{i}} values,
  given in red, are produced by choosing the maximum value of lambda
  that ensures all positive and non-zero \math{x\sp{'}\sb{i}}. These
  values become the new initial values of \math{y\sb{i}} for the next
  iteration cycle.}
\label{fig:lambda}
\end{figure}

To do so, the difference between the initial and final values given by
the Lagrangian calculation, \math{\Delta\sb{i} = x\sb{i} - y\sb{i}},
we will call the total distance for each species. To ensure that all
improved mole numbers are positive, we introduce a new value,
\math{\lambda}, that defines the fraction of the total distance as
\math{\lambda\Delta\sb{i}} (see Figure \ref{fig:lambda}).

The computed changes, \math{\lambda\Delta\sb{i}}, are considered to be
{\em directional numbers} indicating the preferred direction of
descent the system moves to. Other than providing all positive mole
numbers, we determine the value \math{\lambda} so that the Gibbs
energy of the system must decrease, i.e., the minimum point is not
passed (see Equation \ref{direc-deriv}).

At each Lagrangian iteration cycle we start with the initial positive
values, \math{y\sb{i}} and we get the next set of improved values
\math{x\sb{i}} given as:

\begin{eqnarray}
x_i = y_i + \Delta_i \, .
\label{xi-lambda}
\end{eqnarray}

\noindent Since we do not want any \math{x\sb{i}} to be negative, the
variable \math{\lambda} performs a small correction:

\begin{eqnarray}
x_i^{'} = y_i + \lambda\Delta_i \, .
\label{xi_corr}
\end{eqnarray}

\noindent \math{\lambda} takes values between 0 and 1, where value
of {\em zero} implies no step is taken from the iteration's original
input, \math{y\sb{i}}, and {\em one} implies that the full Lagrangian
distance is travelled, \math{\Delta_i}. We now rewrite Equation
(\ref{f}) using Equation (\ref{xi_corr}) as:

\begin{eqnarray}
f_{i}(X^{'}) = x_{i}^{'}\Big(\frac{g_{i}^{0}(T)}{RT} + \ln P +\ln
\frac{x_{i}^{'}}{\bar{x}^{'}}\Big) ,
\label{f_X}
\end{eqnarray}

\noindent which can be written in the form:

\begin{eqnarray}
f_i(\lambda) = (y_i+\lambda\Delta_i)\Big(\frac{g_{i}^0(T)}{RT} + \ln P
+\ln \frac{y_{i}+
  \lambda\Delta_i}{{\bar{y}}+\lambda{\bar{\Delta}}}\Big) ,
\label{f_i_lambda}
\end{eqnarray}

\noindent where \math{\bar{\Delta}} = \math{\bar{y} -
  \bar{x}}. Summing over \math{i}, we get a new function,
\math{F(\lambda)}:

\begin{eqnarray}
F(\lambda) = \sum_i\,(y_i+\lambda\Delta_i)\Big(\frac{g_{i}^0(T)}{RT} +
\ln P +\ln \frac{y_{i}+
  \lambda\Delta_i}{{\bar{y}}+\lambda{\bar{\Delta}}}\Big) .
\label{Flambda}
\end{eqnarray}

\begin{figure*}[h]
\centering
\includegraphics[width=15.5cm, trim=22 100 27 110,
clip=true]{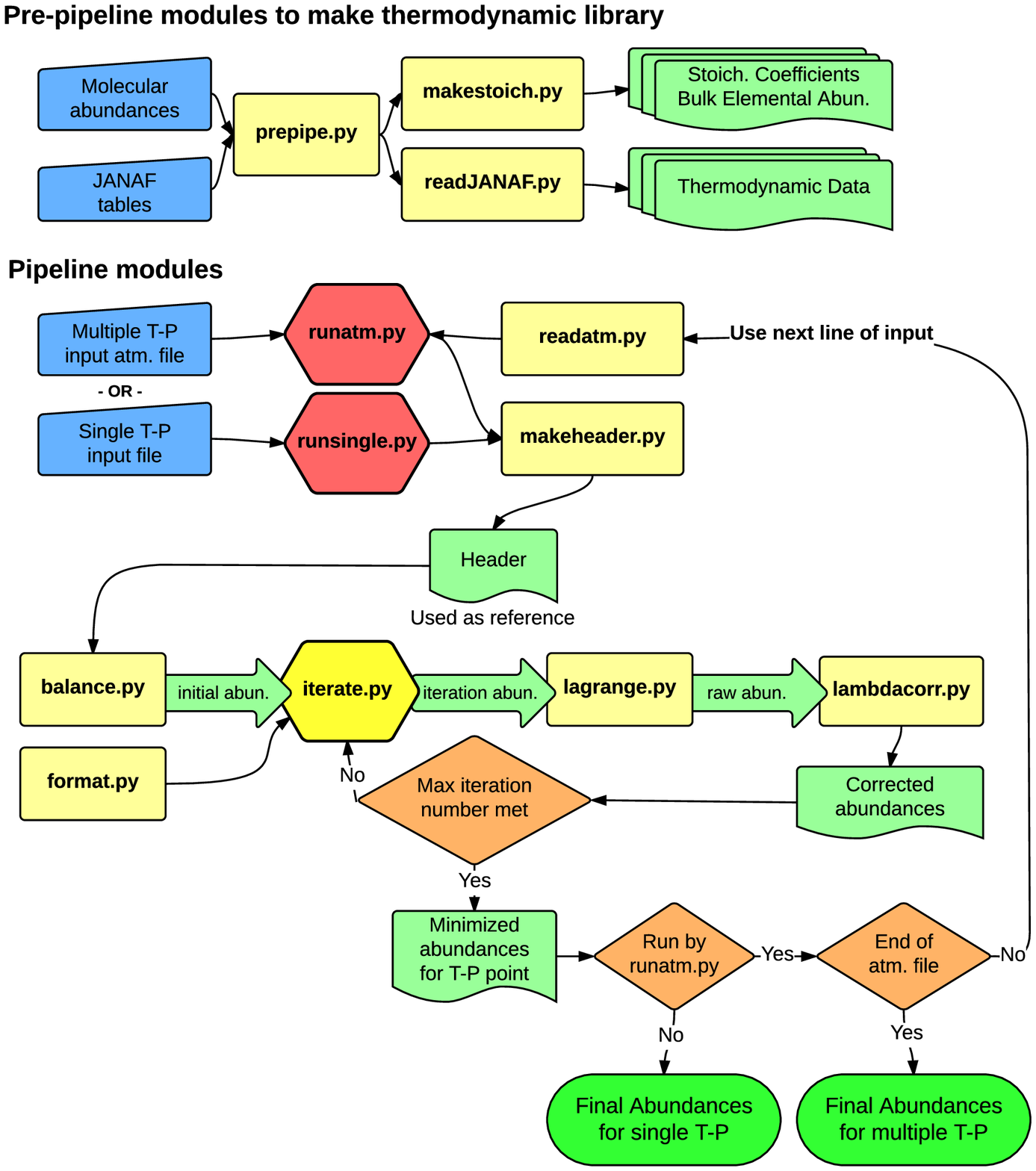}
\includegraphics[width=14.5cm, trim=22 340 27 320,
clip=true]{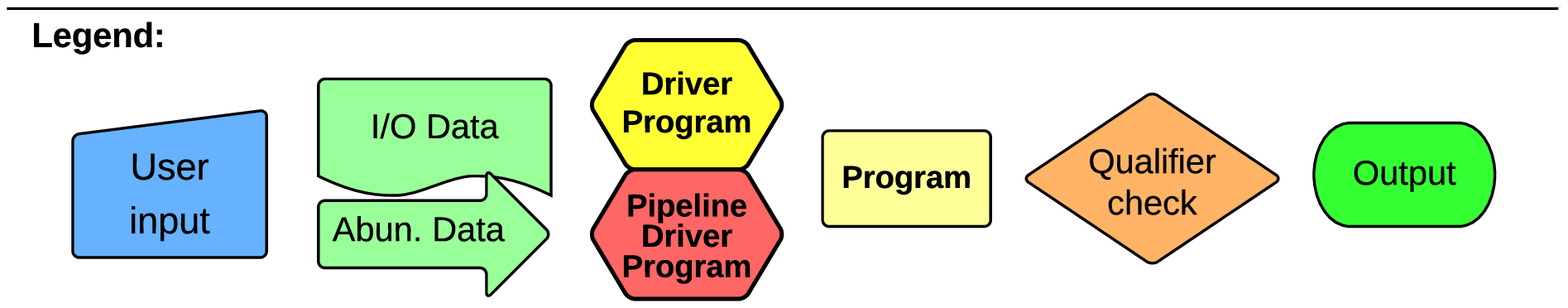}
\caption{Layout of the TEA pre-pipeline and pipeline modules.  The
  modules have one of three roles: scientific calculation, file or
  data structure support, or execution of the calculation programs
  over temperature and pressure points in an iterative manner. In
  addition to the modules shown, TEA has three supporting modules:
  readconfig.py, makeatm.py, and plotTEA.py. All modules are described
  in the text.}
\label{fig:TEAflow}
\end{figure*}

\comment{
We choose \math{\lambda} values using sampling that will ensure
effective exploration of the [0, 1] range. Half of the range is
sampled exponentially, and the other half linearly, totalling 150
points. The exponential sampling is chosen to prevent steps that can
cross the negative threshold for \math{\Delta\sb{i}} (where mole
numbers become negative). For lower temperatures, this threshold has
proven to be very low, thus it requires the smallest step in
\math{\lambda} space. Increasing the number of points in
\math{\lambda} space in certain cases (very low temperatures) can
provide better numerical precision.
}

Thus, to ensure that the new corrected values \math{x\sb{i}\sp{'}} are
all positive, the distance travelled will be limited to fractional
amounts defined by \math{\lambda\Delta\sb{i}}, using the largest
possible value of \math{\lambda} that satisfies the conditions:

{
\begin{enumerate}
\setlength\itemsep{0ex}
\setlength\topsep{0ex}
\setlength\partopsep{0ex}
\setlength\parsep{0ex}

\item The function called the {\em directional derivative} is defined
  and exists:

\begin{eqnarray}
\frac{dF(\lambda)}{d\lambda} = \sum_{i=1}^n \Delta_i
\Big[\frac{g_{i}^0(T)}{RT} + \ln P + \ln \frac{y_{i} +
    \lambda\Delta_i}{\bar{y} + \lambda\bar{\Delta}}\Big] \, .
\label{direc-deriv}
\end{eqnarray}

\item The directional derivative does not become positive (the minimum
  point is not passed).

\end{enumerate}
}

Every new iteration starts with a different set of \math{y\sb{i}},
thus changing the convex function \math{F(\lambda)}, Equation
\ref{Flambda}, and producing a new minimum. This yields to a new
\math{\lambda} value.  \math{\lambda} will be found to approach unity
after some number of iterations. Unity in \math{\lambda} indicates the
solution is near.

We repeat the Lagrangian method and the lambda correction until a
pre-defined maximum number of iterations is met. The final abundances
are given as fractional abundances (mole mixing fractions), i.e., the
ratio of each species' mole numbers to the total sum of mole numbers
of all species in the mixture.

\section{Code Structure}
\label{sec:struct}

The TEA code is written entirely in Python and uses the Python
packages NumPy ({\tt http://numpy.org/}) and ({\tt
  http://www.scipy.org/}) along with SymPy, an external linear
equation solver ({\tt http://sympy.org/}).

The code is divided into two parts: the pre-pipeline that makes the
thermochemical data library and stoichiometric tables, and the
pipeline that performs abundance calculations.  Given elemental
abundances, TEA calculates molecular abundances for a particular
temperature and pressure or a list of temperature-pressure
pairs. Documentation is provided in the TEA User Manual (Bowman and
Blecic) and the TEA Code Description (Blecic and Bowman) that
accompany the code. Figure \ref{fig:TEAflow} shows the layout of the
TEA program's flow.  Its modules are:

\begin{figure*}[!t]
\hspace{100pt}
\includegraphics[height=5cm]{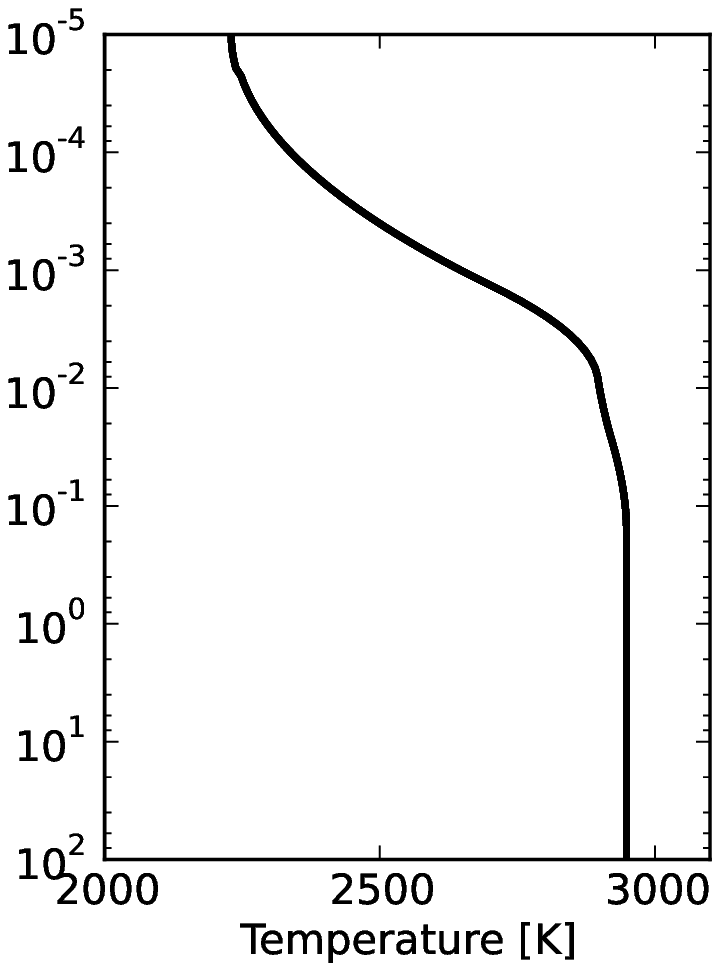}
\includegraphics[height=5cm]{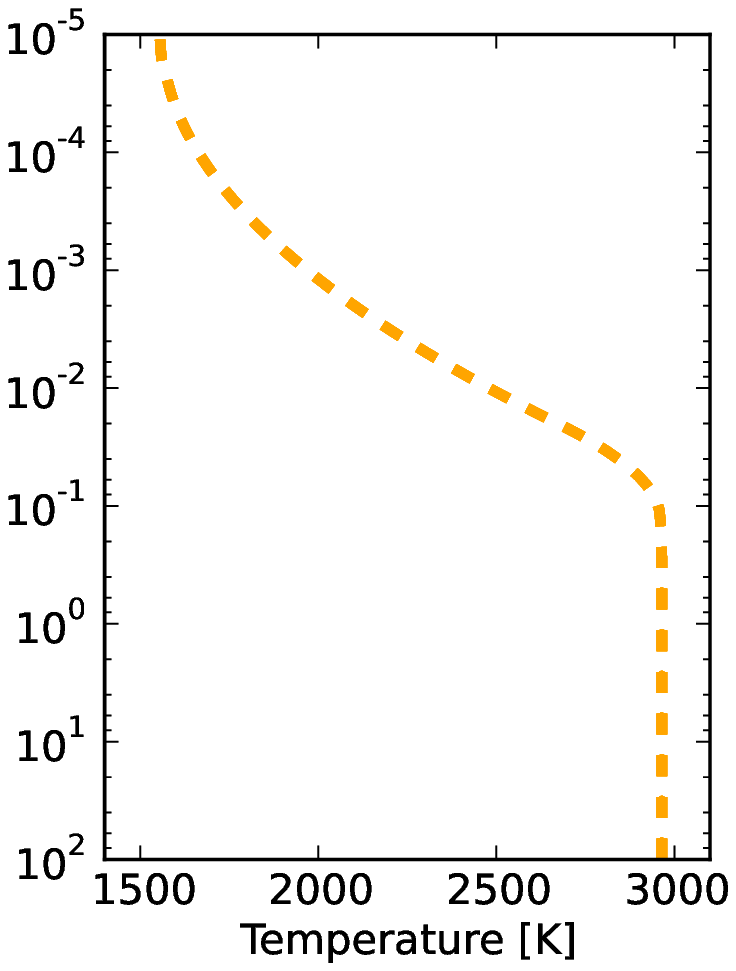}
\includegraphics[height=5cm]{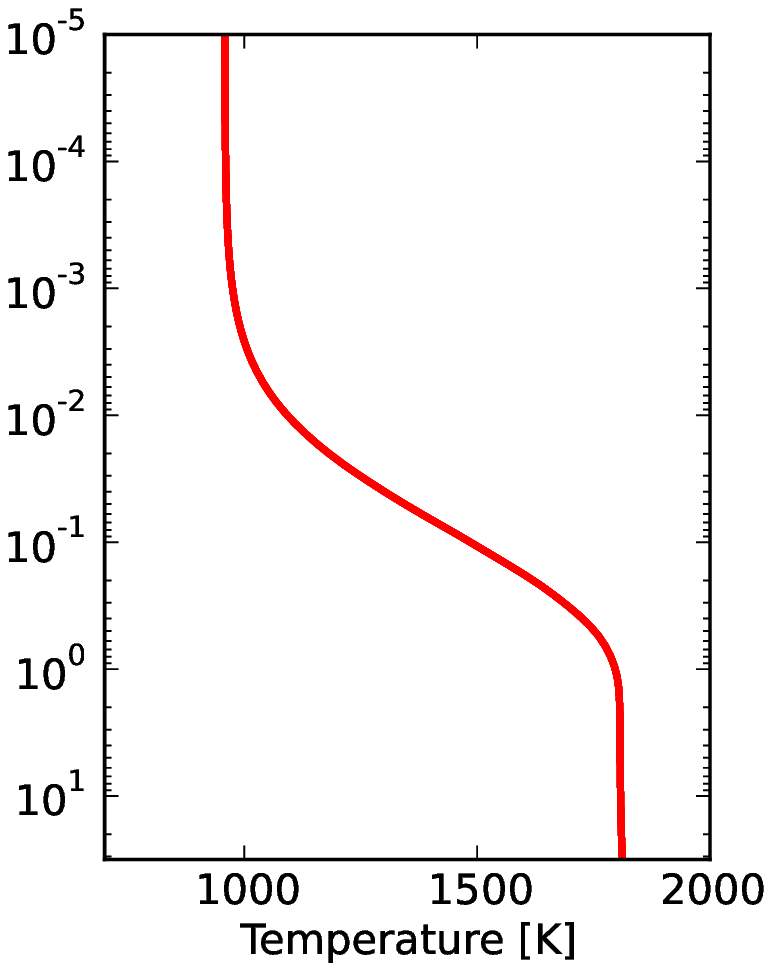}
\figcaption{\label{fig:profiles} The left and middle panels show the
  O-rich and C-rich temperature and pressure (\math{T-P}) profile of
  WASP-12b from \citet{StevensonEtal2014-WASP12b} with C/O = 0.5 and
  C/O = 1.2 respectively. The right panel shows the \math{T-P} profile
  of WASP-43b from \citet{StevensonEtal2014-PhaseCurve-WASP43b} with
  solar metallicity.}  \includegraphics[width=0.50\linewidth,
  clip]{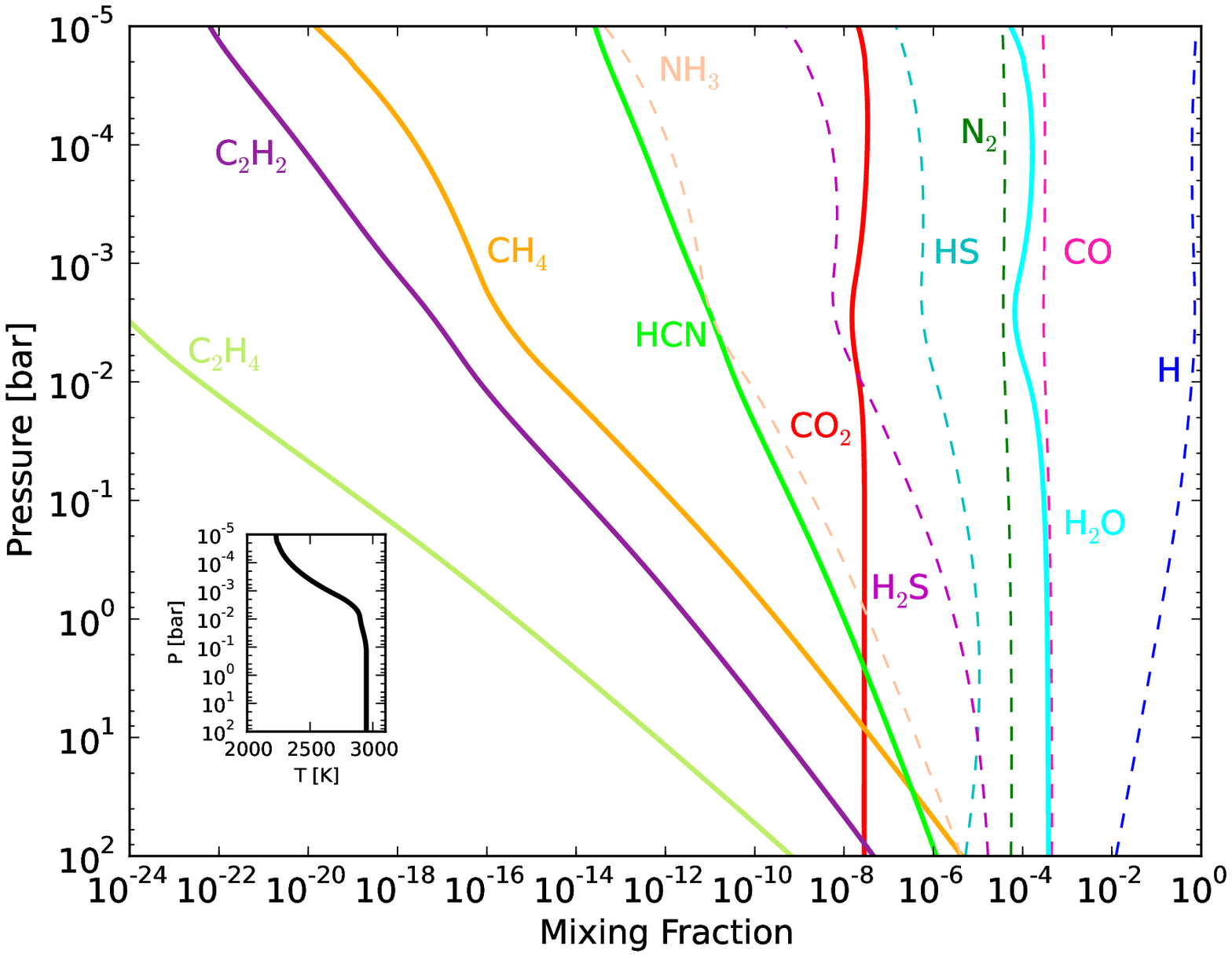}
\includegraphics[width=0.50\linewidth,
  clip]{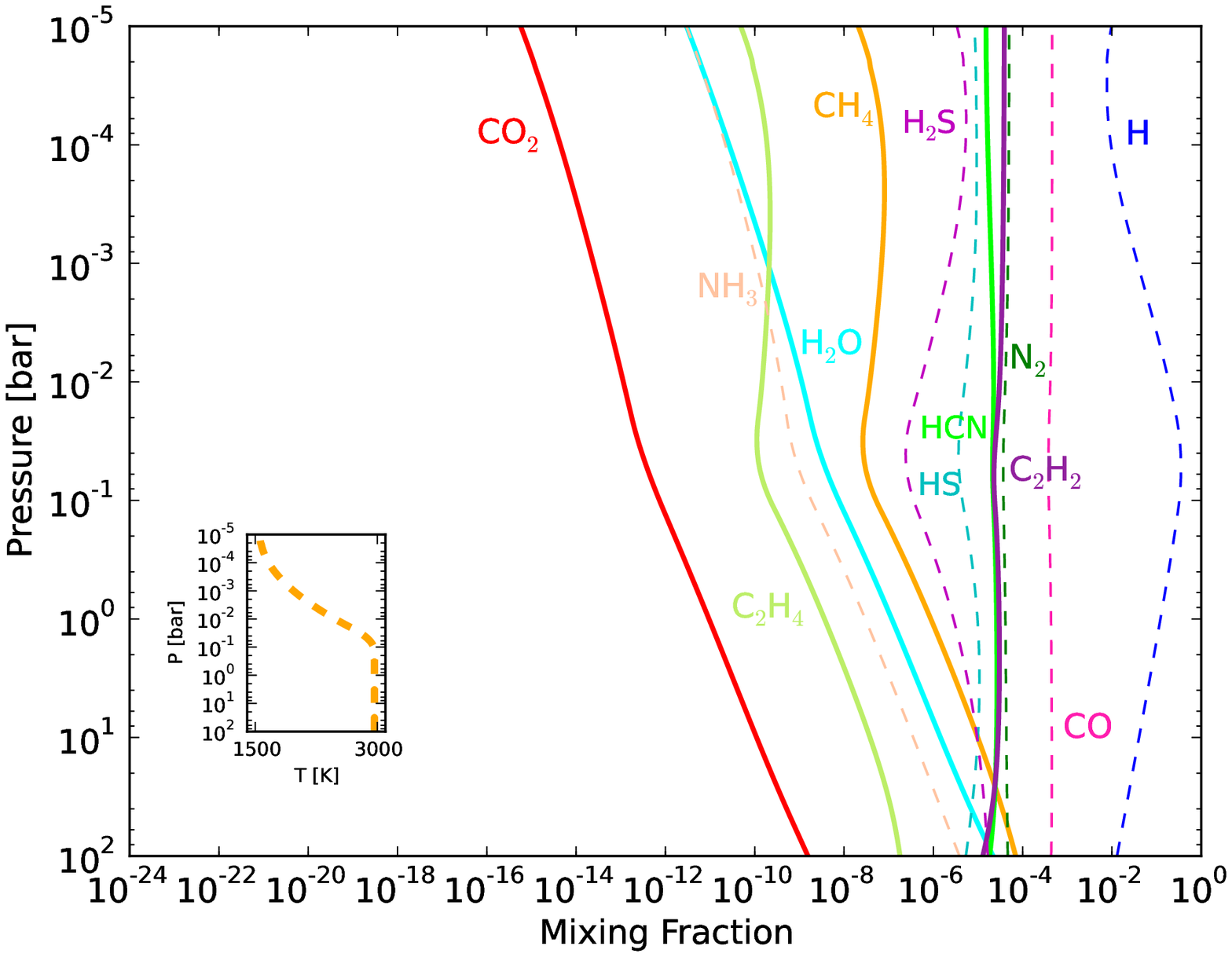} \figcaption{\label{fig:WASP-12b}
  Comparison between vertical thermochemical equilibrium distributions
  for WASP-12b O-rich, left panel, and WASP-12b C-rich, right panel,
  elemental abundance profile. The inset plots are the \math{T-P}
  profiles shown in Figure \ref{fig:profiles}, left and middle
  panels.}
\end{figure*}

{
\begin{enumerate}
\setlength\itemsep{0ex}
\setlength\topsep{0ex}
\setlength\partopsep{0ex}
\setlength\parsep{0ex}

   \item {\bf prepipe.py}: Runs the {\tt readJANAF.py} and {\tt
     makestoich.py} modules and provides their common setup.
   \item {\bf readJANAF.py}: Extracts relevant from all available
     NIST-JANAF Thermochemical Tables and writes ASCII files.
   \item {\bf makestoich.py}: Reads the chemical formula to obtain
     species names and their stoichiometric coefficients from each
     JANAF file, and elemental solar abundances from an ASCII file
     based on \citet{AsplundEtal2009-SunAbundances} Table 1. The code
     produces an output file containing species, stoichiometric
     coefficients, and abundances.
   \item {\bf runsingle.py}: Runs TEA for a single \math{T, P} pair.
   \item {\bf runatm.py}: Runs TEA over a pre-atmosphere file
     containing a list of \math{T, P} pairs.
   \item {\bf readatm.py}: Reads the pre-atmospheric file with
     multiple \math{T, P} pairs.
   \item {\bf makeheader.py}: Combines the stoichiometric information,
     Gibbs free energy per species at specific temperatures, and the
     user input to create a single file with relevant chemical
     informations further used by the pipeline. 
   \item {\bf balance.py}: Uses species and stoichiometric
     information to establish viable, mass-balanced, initial mole
     numbers.
   \item {\bf format.py}: Auxiliary program that manages input/output
     operations in each piece of the pipeline.
   \item {\bf lagrange.py}: Uses data from the most recent
     iteration's corrected mole numbers and implements the Lagrangian
     method for minimization.  Produces output with raw, non-corrected
     mole numbers for each species (values are temporarily allowed to
     be negative).
   \item {\bf lambdacorr.py}: Takes non-corrected mole numbers and
     implements lambda correction to obtain only valid, positive
     numbers of moles.  Output is the corrected mole
     numbers for each species.
   \item {\bf iterate.py}: Driver program that repeats {\em
     lagrange.py} and {\em lambdacorr.py} until a pre-defined maximum
     number of iterations is met.
   \item {\bf readconfig.py}: Reads TEA configuration file.
   \item {\bf makeatm.py}: Makes pre-atmospheric file for a multiple
     \math{T, P} run.
   \item {\bf plotTEA.py}: Plots TEA output, the atmospheric file with
     final mole-fraction abundances.
\end{enumerate}
}

\section{Application to hot-Jupiter atmospheres}
\label{sec:applic}
In this section, we illustrate several applications of the TEA
code. We produced molecular abundances profiles for models of
hot-Jupiter planetary atmospheres, given their temperature-pressure
profiles.

The temperature and pressure (\math{T-P}) profiles adopted for our
thermochemical calculations are shown in Figure
\ref{fig:profiles}. The left and middle panel show the \math{T-P}
profiles of WASP-12b from \citet{StevensonEtal2014-WASP12b} with the
C/O ratio of 0.5 and 1.2, respectively.  The right panel shows the
thermal profile of WASP-43b from
\citet{StevensonEtal2014-PhaseCurve-WASP43b} with solar
metallicity. These profiles are chosen for their relevance to
atmospheric conditions at secondary eclipses.

We chose elemental-abundance profiles with C/O \math{>} 1 and C/O
\math{<} 1 and three profiles with solar, 10 times solar, and 50 times
solar elemental abundances to show the influence of the C/O ratio and
metallicity on the chemistry and composition of extrasolar giant
planets.

We adopt \citet{AsplundEtal2009-SunAbundances} photospheric solar
abundances as our baseline. To change the elemental abundance profile,
set them to a certain C/O ratio, or enhance metallicity, we use our
Python routine, {\tt makeAbun.py}. This routines is the part of the
BART project and it is available to the community via {\tt Github.com}
under an open-source licence ({\tt
  https://github.com/joeharr4/BART}). For different metallicities, the
routine multiples the elemental abundances of all species except for
hydrogen and helium, preserving the ratio of major atomic species like
C, N, and O.  \comment{ To get a certain C/O ratio \math{>} 1, we
  swapped the elemental abundances of carbon and oxygen and then
  decrease the carbon elemental abundance to get the desired ratio.  }

We chose to run the models for all plausible, spectroscopically active
species in the infrared relevant for hot-Jupiter atmospheres: H\sb{2},
CO, CO\sb{2}, CH\sb{4}, H\sb{2}O, HCN, C\sb{2}H\sb{2}, C\sb{2}H\sb{4},
N\sb{2}, NH\sb{3}, HS, and H\sb{2}S. Our input species are: H, He, C,
N, O, S.


\begin{figure}[t!]
    \centering \includegraphics[width=0.99\linewidth,
      clip]{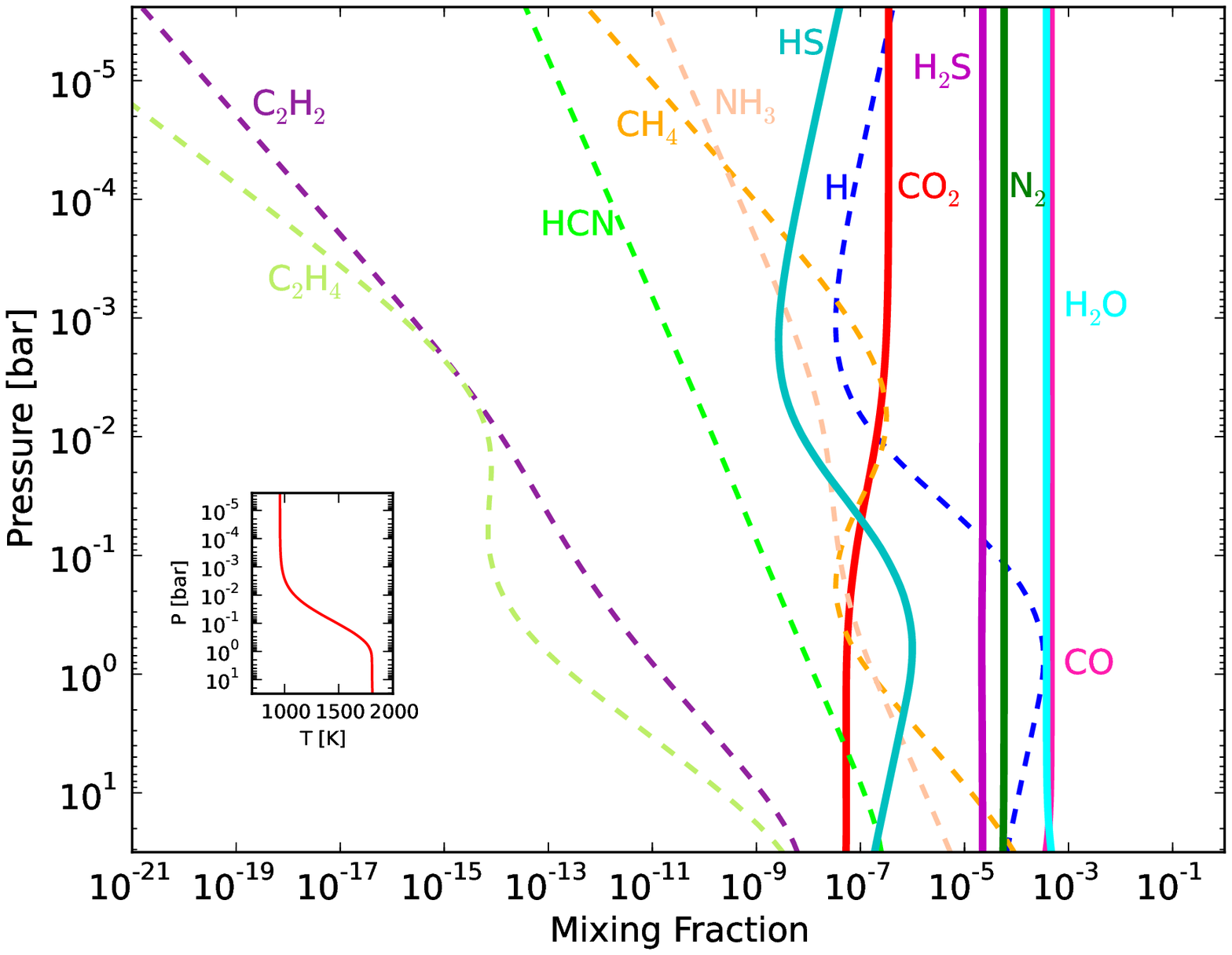}
    \includegraphics[width=0.99\linewidth,
      clip]{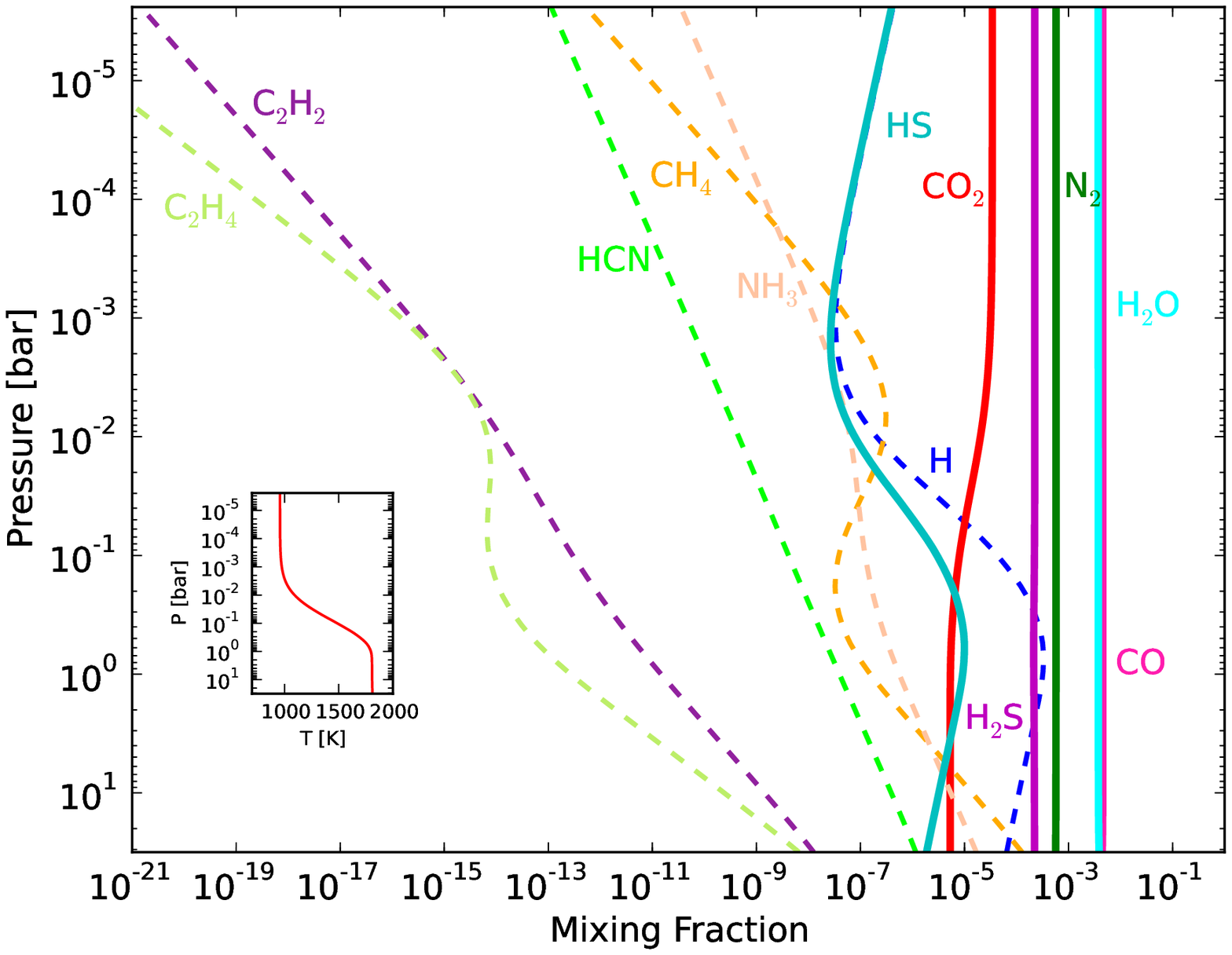}
    \includegraphics[width=0.99\linewidth,
      clip]{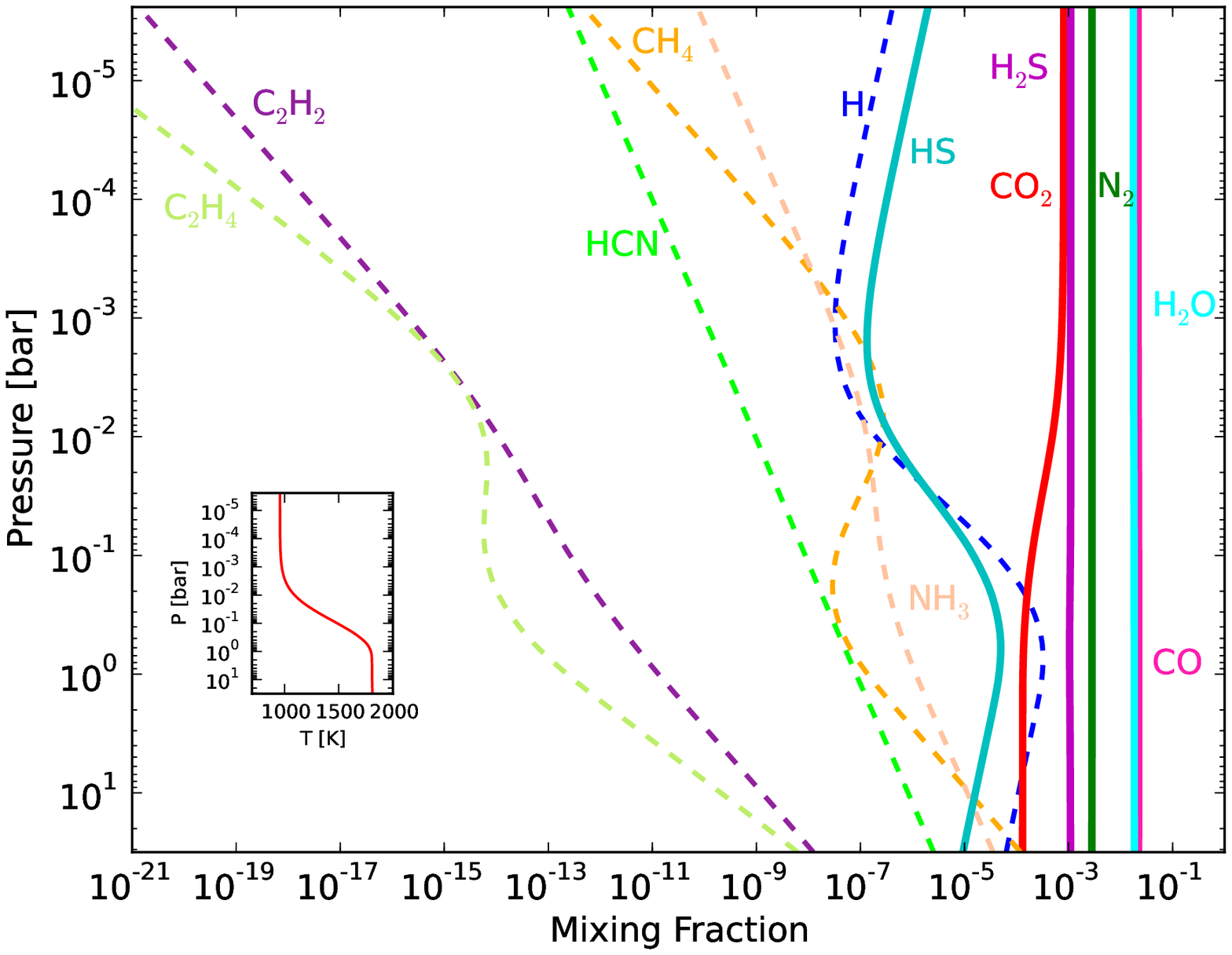}
    \figcaption{\label{fig:WASP-43b-metal} Thermochemical equilibrium
      vertical distributions for different metallicities of WASP-43b
      assuming the \math{T-P} profile in Figure \ref{fig:profiles},
      right panel (profile given in inset). Three metallicity cases
      with \math{\zeta}= 1, 10, and 50 are shown from the top to the
      bottom.}
\end{figure}

Figure \ref{fig:WASP-12b} shows results for WASP-12b. Each \math{T-P}
profile is sampled 100 times uniformly in log-pressure space. Figure
\ref{fig:WASP-43b-metal} shows the TEA runs for WASP-43b with
different metallicities. This \math{T-P} profile is sampled 90 times in
uniformly log-pressure space.

As expected, Figure \ref{fig:WASP-12b} shows that H\sb{2}O, CH\sb{4},
CO, CO\sb{2}, C\sb{2}H\sb{2}, C\sb{2}H\sb{4}, and HCN are under the
strong influence of the atmospheric C/O ratio in hot Jupiters
\citep[e.g., ][]{Lodders02, Seager05, FortneyEtal2005apjlhjmodels,
  MadhusudhanEtal2011natWASP12batm, MadhusudhanEtal2011-Cplanets,
  Madhusudhan2012-COratio, MadhusudhanSeager2011-GJ436b,
  MosesEtal2013-COratio}. These species are plotted in solid lines,
while species with only small influence from the C/O ratio are plotted
as dashed lines.

The results also show, as expected, that CO is a major atmospheric
species on hot Jupiters for all C/O ratios and metallicities (Figures
\ref{fig:WASP-12b} and \ref{fig:WASP-43b-metal}), because CO is
chemically favored over H\sb{2}O. Other oxygen-bearing molecules like
H\sb{2}O and CO\sb{2} are more abundant when C/O\math{<}1, while
CH\sb{4}, C\sb{2}H\sb{2}, and C\sb{2}H\sb{4} become significant
species when C/O\math{>}1. Species like N\sb{2} and NH\sb{3} that do
not contain carbon or oxygen are much less affected by the C/O ratio.

H\sb{2}O is abundant in hot-Jupiter atmospheres \citep[e.g.,
][]{BurrowsSharp1999apjchemeq, Lodders02, HubenyBurrows2007,
  SharpBurrows2007Apjopacities} due to the large solar abundances of
oxygen and hydrogen. Even disequilibrium processes like photochemistry
cannot deplete its abundance. Photochemical models by
\citet{MosesEtal2011-diseq} and \citet{LineEtal2010ApJHD189733b,
  LineEtal2011-kinetics} predict that water will be recycled in
hot-Jupiter atmospheres, keeping H\sb{2}O abundances close to
thermochemical equilibrium values. A low water abundance seems to
occur only in atmospheres with a C/O\math{>}1.

CO\sb{2}, although present in hot-Jupiter atmospheres and
spectroscopically important, is not a major constituent, and it
becomes even less abundant when C/O\math{>}1. Although photochemistry
can greatly enhance the HCN, C\sb{2}H\sb{2}, and C\sb{2}H\sb{4}
abundances \citep{MosesEtal2013-COratio}, we also see that with
C/O\math{>}1, they are the most abundant constituents.

In Figure \ref{fig:WASP-43b-metal}, the species strongly influenced by
metallicity are again plotted as solid lines. In general, we see, as
expected \citep[e.g., ][]{LineEtal2011-kinetics, Lodders02,
  Venot2014-metallicity}, that the shapes of the vertical
distributions are mostly preserved for all metallicities. However, the
thermochemical mixing ratio of CO\sb{2}, CO, H\sb{2}O, N\sb{2}, HS,
and H\sb{2}S vary by several orders of magnitude over the range of
metallicities, while CH\sb{4} and hydrocarbons change very little.

When the metallicity changes from 1 to 50, the abundance of CO\sb{2}
experiences the most dramatic change. It increases by a factor of
1000, confirming it as the best probe of planetary metallicity
\citep{Lodders02, Zahnle09-SulfurPhotoch}.  CO\sb{2} abundance is the
quadratic function of metallicity \citep{Venot2014-metallicity}, while
CO, H\sb{2}O, HS, H\sb{2}S, and N\sb{2} abundances, for species that
either contain one metal atom or are the major reservoirs of carbon
and nitrogen, increase linearly with metallicity \citep{
  Visscher2006}. For this metallicity range, the CO, H\sb{2}O, HS,
H\sb{2}S, and N\sb{2} abundances change by a factor of 100, while
NH\sb{3}, CH\sb{4}, C\sb{2}H\sb{2}, C\sb{2}H\sb{4}, and HCN change by
a factor of 10 or less.

\section{Comparison to other methods}
\label{sec:validity}

\begin{figure*}[!t]
\centering \includegraphics[height=6.8cm, clip = True, trim=0.75cm
  0.1cm 0.1cm 0cm]{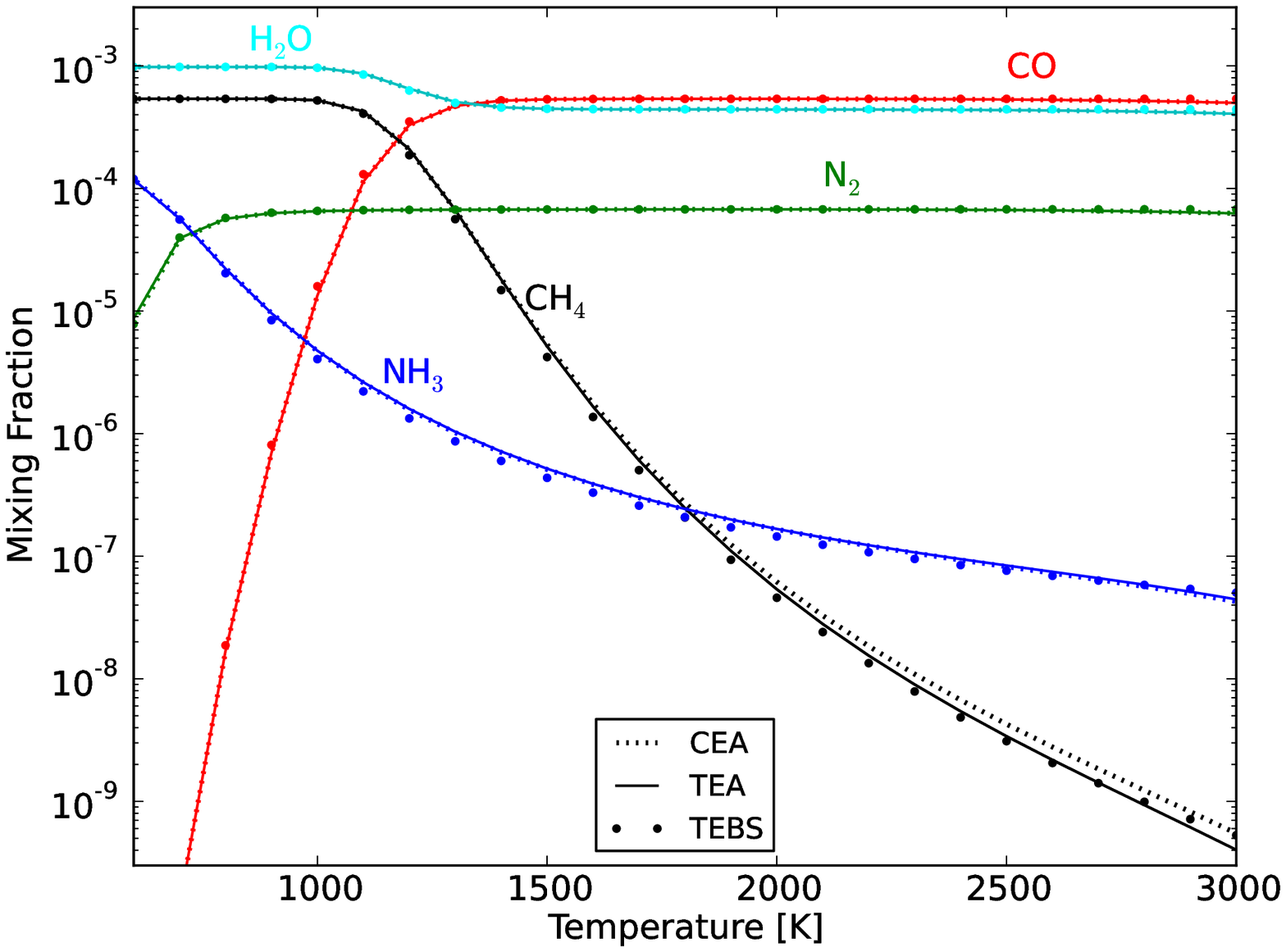}
\includegraphics[height=6.8cm, clip = True, trim=0.75cm 0.1cm 0.1cm
  0cm]{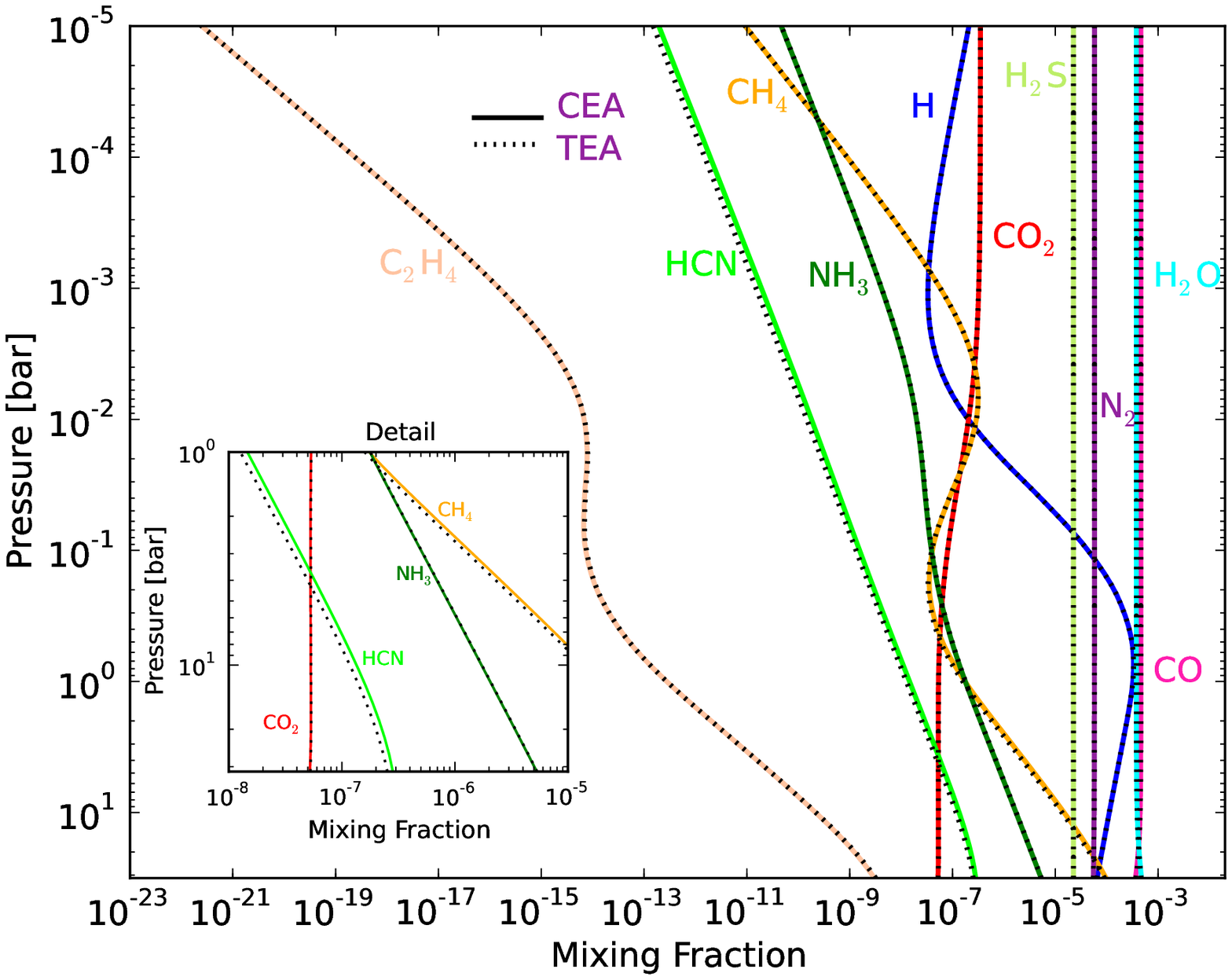}
\caption{{\bf Left:}\label{fig:CEATEATEBS} Comparison TEA, CEA and
  TEBS. TEBS is an analytic method, while CEA and TEA are numerical
  methods. We show the major spectroscopically-active species in the
  infrared that can be produced by all three methods. We run the codes
  for the same range of temperatures and the pressure of \math{P} = 1
  bar. Each species in each method is plotted with a different line
  style, but with the same color. The TEBS final abundances are
  plotted as dots, CEA as dashed lines, while TEA is plotted as solid
  lines. {\bf Right:} Comparison of the TEA results with the results
  from CEA. CEA and TEA are both numerical methods that use Gibbs free
  energy minimization method with similar optimization scheme. We show
  the most plausible and most abundant spectroscopically-active
  species in the infrared expected to be present in hot-Jupiter
  atmospheres, that all codes can cover. In the inset plot, we show a
  detail (zoom-in part), pointing out species lines
  that do not overlap.  The \math{T-P} profile used for this run is
  given in the right panel of Figure \ref{fig:profiles}. Tables
  \ref{table:CEA-TEA-left} and \ref{table:CEA-TEA-right} list
  differences between the final abundances for random three \math{T,
    P} points chosen from each run.}
\end{figure*}

To test the validity of our code, we performed 4 different tests.  We
compared the output of TEA with the example from
\citet{WhiteJohnsonDantzig1958JGibbs} using their thermodynamic data.
We also compared the TEA output with the output of our TEBS
(Thermochemical Equilibrium by \citeauthor{BurrowsSharp1999apjchemeq})
code that implements the \citet{BurrowsSharp1999apjchemeq} analytical
method for calculating the abundances of five major molecular species
present in hot-Jupiter atmospheres (CO, CH4, H2O, N2, NH3). As another
comparison, we used the free thermochemical equilibrium code CEA
(Chemical Equilibrium with Applications, available from NASA Glenn
Research Center at {\tt http://www.grc.nasa.gov/WWW/CEAWeb/}). This
code uses the Newton-Raphson descent method within the Lagrange
optimization scheme to solve for chemical abundances. Their approach
is described by \citet{GordonMcBride:1994, McBrideGordon:1996}, and
\citet{ZeleznikGordon:1960, ZeleznikGordon:1968}. The thermodynamic
data included in the CEA code are partially from the JANAF tables
\citep{ChaseEtal1986bookJANAFtables} that we used in our TEA code, but
also from numerous other sources \citep[e.g.,][]{Cox1982-CEAdata,
  Gurvich1989-CEAdata, McBrideGordonReno:1993a}. Lastly, we derived
CEA free energies and used them as input to TEA, to compare the CEA
and TEA outputs.

Our first comparison was done using the example from
\citet{WhiteJohnsonDantzig1958JGibbs}. We determined the composition
of the gaseous species arising from the combustion of a mixture of
hydrazine, N\sb{2}H\sb{4}, and oxygen, O\sb{2}, at \math{T} = 3500 K
and the pressure of 750 psi = 51.034 atm. We used the free-energy
functions and \math{b\sb{j}} values (total number of moles of element
j originally present in the mixture) from their Table 1. We reproduced
their abundances, Table \ref{table:White-TEA}, with slightly higher
precision probably due to our use of double precision.

\begin{table}[!h]
\caption{\label{table:White-TEA} Comparison
  \citeauthor{WhiteJohnsonDantzig1958JGibbs} {\em vs.} TEA}
\atabon\strut\hfill\begin{tabular}{lcccc} \hline \hline Species &
\math{\frac{g\sb{i}\sp{0}(T)}{RT}} &
\citeauthor{WhiteJohnsonDantzig1958JGibbs} & TEA & Difference \\ & &
abundances & abundances & \\ \hline H & -10.021 & 0.040668
& 0.04065477 & -0.00001323 \\ H\sb{2} & -21.096 & 0.147730 &
0.14771009 & -0.00001991 \\ H\sb{2}O & -37.986 & 0.783153 & 0.78318741
& 0.00003441 \\ N & -9.846 & 0.001414 & 0.00141385 & -0.00000015
\\ N\sb{2} & -28.653 & 0.485247 & 0.48524791 & 0.00000091 \\ NH &
-18.918 & 0.000693 & 0.00069312 & 0.00000012 \\ NO & -28.032 &
0.027399 & 0.02739720 & -0.00000180 \\ O & -14.640 & 0.017947 &
0.01794123 & -0.00000577 \\ O\sb{2} & -30.594 & 0.037314 & 0.03730853
& -0.00000547 \\ OH & -26.111 & 0.096872 & 0.09685710 & 0.00001490
\\ \hline
\end{tabular}\hfill\strut\ataboff
\end{table}


\begin{figure*}[!t]
\centering \includegraphics[height=6.8cm, clip = True, trim=0.75cm
  0.1cm 0.1cm 0cm]{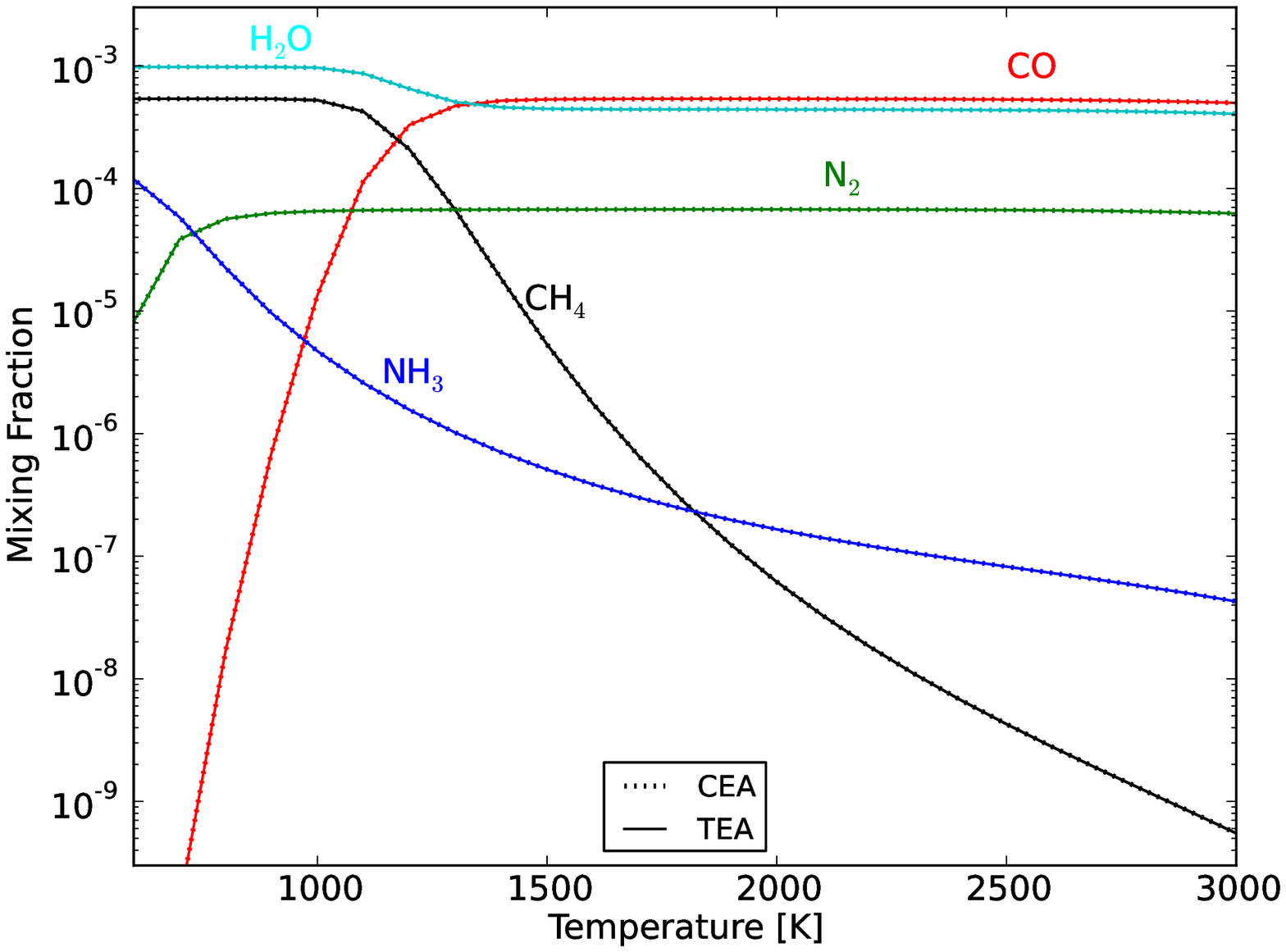}
\includegraphics[height=6.8cm, clip = True, trim=0.75cm 0.1cm 0.1cm
  0cm]{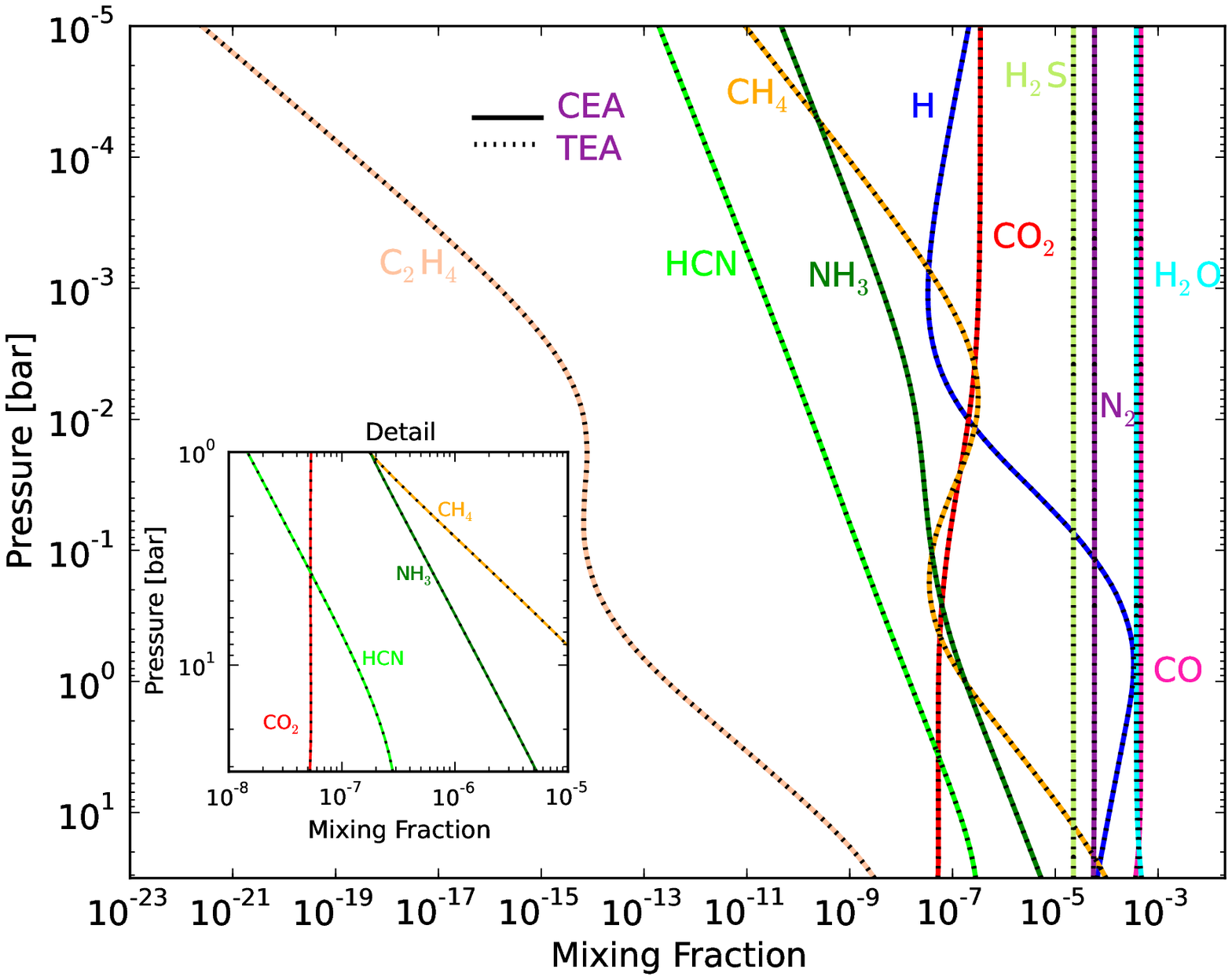}
\caption{{\bf Left:}\label{fig:CEATEA} Comparison TEA and CEA using
  CEA thermodynamic data provided in their {\tt thermo.inp} file. The
  comparison is done for the same conditions as in Figure
  \ref{fig:CEATEATEBS}. Tables \ref{table:CEA-TEA-left} and
  \ref{table:CEA-TEA-right} list differences between the final
  abundances for random three \math{T, P} points chosen from each
  run.}
\end{figure*}

\begin{table*}[!t]
\caption{\label{table:CEA-TEA-left} Differences CEA {\em vs.} TEA,
  Figures \ref{fig:CEATEATEBS} and \ref{fig:CEATEA}, Left Panels}
\atabon\strut\hfill\begin{tabular}{lcccccc} \hline \hline Pressure &
Temp & \multicolumn{5}{c}{Species} \\ (bar) & (K) & CO & CH4 & H2O &
N2 & NH3 \\ \hline \multicolumn{7}{c}{\bf CEA free energies} \\ \hline
1.0000e+00 & 2500.00 & -33.80559930 & -34.91655970 & -40.12912409 &
-27.71757996 & -32.70878374 \\ 1.0000e+00 & 2700.00 & -33.69182758 &
-35.33414843 & -39.64186320 & -27.99507610 & -33.05270703
\\ 1.0000e+00 & 2900.00 & -33.61649725 & -35.76252669 & -39.25604592 &
-28.25692510 & -33.39666924 \\ \hline \multicolumn{7}{c}{\bf TEA
  (JANAF) free energies} \\ \hline 1.0000e+00 & 2500.00 & -33.80793700
& -34.70780992 & -40.12426098 & -27.72037451 & -32.73695542
\\ 1.0000e+00 & 2700.00 & -33.69214791 & -35.08662806 & -39.63255004 &
-27.99496246 & -33.08302621 \\ 1.0000e+00 & 2900.00 & -33.61466712 &
-35.47533692 & -39.24191944 & -28.25439783 & -33.42896561 \\ \hline
\multicolumn{7}{c}{\bf CEA final abundances} \\ \hline 1.0000e+00 &
2500.00 & 5.3129e-04 & 4.2666e-09 & 4.3546e-04 & 6.6686e-05 &
8.2252e-08 \\ 1.0000e+00 & 2700.00 & 5.2311e-04 & 1.8387e-09 &
4.2855e-04 & 6.5661e-05 & 6.4332e-08 \\ 1.0000e+00 & 2900.00 &
5.0876e-04 & 8.2340e-10 & 4.1586e-04 & 6.3844e-05 & 4.9466e-08
\\ \hline \multicolumn{7}{c}{\bf TEA final abundances using CEA free
  energies} \\ \hline 1.0000e+00 & 2500.00 & 5.3129e-04 & 4.2665e-09 &
4.3547e-04 & 6.6685e-05 & 8.2251e-08 \\ 1.0000e+00 & 2700.00 &
5.2311e-04 & 1.8387e-09 & 4.2856e-04 & 6.5661e-05 & 6.4332e-08
\\ 1.0000e+00 & 2900.00 & 5.0876e-04 & 8.2339e-10 & 4.1586e-04 &
6.3844e-05 & 4.9466e-08 \\ \hline \multicolumn{7}{c}{\bf TEA final
  abundances using JANAF free energies} \\ \hline 1.0000e+00 & 2500.00
& 5.3129e-04 & 3.3976e-09 & 4.3547e-04 & 6.6685e-05 & 8.3987e-08
\\ 1.0000e+00 & 2700.00 & 5.2312e-04 & 1.4194e-09 & 4.2856e-04 &
6.5661e-05 & 6.6260e-08 \\ 1.0000e+00 & 2900.00 & 5.0878e-04 &
6.1471e-10 & 4.1586e-04 & 6.3845e-05 & 5.1339e-08 \\ \hline
\end{tabular}\hfill\strut\ataboff
\end{table*}

\begin{table*}[!t]
\caption{\label{table:CEA-TEA-right} Differences CEA {\em vs.} TEA,
  Figures \ref{fig:CEATEATEBS} and \ref{fig:CEATEA}, Right Panels}
\atabon\strut\hfill\begin{tabular}{lcccccccc} \hline \hline Pressure &
Temp & \multicolumn{7}{c}{Species} \\ (bar) & (K) & CO & CO2 & CH4 &
H2O & HCN & NH3 & H2S \\ \hline \multicolumn{9}{c}{\bf CEA free
  energies} \\ \hline 3.8019e-01 & 1719.64 & -34.9307244 & -58.4124145
& -33.595356 & -43.7404858 & -19.8124391 & -31.4721776 & -30.6054338
\\ 1.6596e+00 & 1805.28 & -34.7237371 & -57.3627896 & -33.695587 &
-43.1403205 & -20.4938031 & -31.5886558 & -30.7577826 \\ 2.1878e+01 &
1810.15 & -34.7128505 & -57.3065771 & -33.701803 & -43.1083097 &
-20.5310828 & -31.5955222 & -30.7664570 \\ \hline
\multicolumn{9}{c}{\bf TEA (JANAF) free energies} \\ \hline 3.8019e-01
& 1719.64 & -34.9288052 & -58.4130468 & -33.5180429 & -43.7386156 &
-19.6678405 & -31.4830322 & -30.5760402 \\ 1.6596e+00 & 1805.28 &
-34.7231685 & -57.3648328 & -33.6061366 & -43.1392058 & -20.3573682 &
-31.6020990 & -30.7285988 \\ 2.1878e+01 & 1810.15 & -34.7123566 &
-57.3086978 & -33.6116396 & -43.1072342 & -20.3950903 & -31.6091097 &
-30.7372812 \\ \hline \multicolumn{9}{c}{\bf CEA final abundances}
\\ \hline 3.8019e-01 & 1719.64 & 4.5960e-04 & 5.8035e-08 & 4.9221e-08
& 3.7681e-04 & 5.6243e-09 & 7.9716e-08 & 2.2498e-05 \\ 1.6596e+00 &
1805.28 & 4.5918e-04 & 5.2864e-08 & 4.4665e-07 & 3.7724e-04 &
2.4131e-08 & 2.8864e-07 & 2.2504e-05 \\ 2.1878e+01 & 1810.15 &
4.0264e-04 & 5.3052e-08 & 5.6873e-05 & 4.3396e-04 & 2.3851e-07 &
3.7082e-06 & 2.2520e-05 \\ \hline \multicolumn{9}{c}{\bf TEA final
  abundances using CEA free energies} \\ \hline 3.8019e-01 & 1719.64 &
4.5959e-04 & 5.8035e-08 & 4.9219e-08 & 3.7682e-04 & 5.6240e-09 &
7.9714e-08 & 2.2498e-05 \\ 1.6596e+00 & 1805.28 & 4.5918e-04 &
5.2865e-08 & 4.4667e-07 & 3.7724e-04 & 2.4131e-08 & 2.8864e-07 &
2.2504e-05 \\ 2.1878e+01 & 1810.15 & 4.0263e-04 & 5.3053e-08 &
5.6875e-05 & 4.3396e-04 & 2.3851e-07 & 3.7083e-06 & 2.2519e-05
\\ \hline \multicolumn{9}{c}{\bf TEA final abundances using JANAF free
  energies} \\ \hline 3.8019e-01 & 1719.64 & 4.5959e-04 & 5.8326e-08 &
4.5480e-08 & 3.7681e-04 & 4.8604e-09 & 8.0472e-08 & 2.2497e-05
\\ 1.6596e+00 & 1805.28 & 4.5922e-04 & 5.3200e-08 & 4.0512e-07 &
3.7719e-04 & 2.0937e-08 & 2.9102e-07 & 2.2504e-05 \\ 2.1878e+01 &
1810.15 & 4.0694e-04 & 5.3429e-08 & 5.2592e-05 & 4.2965e-04 &
2.1124e-07 & 3.7386e-06 & 2.2519e-05 \\ \hline
\end{tabular}\hfill\strut\ataboff
\end{table*}

Figure \ref{fig:CEATEATEBS}, left panel, shows the CEA, TEA, and TEBS
runs for the temperatures between 600 and 3000 K, pressure of 1 bar,
and solar abundances. The runs were performed with the input and
output species that all codes contain (H, C, O, N, H\sb{2}, CO,
CH\sb{4}, H\sb{2}O, N\sb{2}, NH\sb{3}). We also run the comparison
just between CEA and TEA, Figure \ref{fig:CEATEATEBS}, right panel,
for the WASP-43b model atmosphere that we described in Section
\ref{sec:applic}.  We used the pressure and temperature profile shown
in Figure \ref{fig:profiles}, right panel, and solar elemental
abundances. The temperatures and pressures range from 958.48 to
1811.89 K and 1.5\math{\times}10\sp{-5} to
3.1623\math{\times}10\sp{1} bar, respectively. We included the same
species as in Section \ref{sec:applic} with the exclusion of the
C\sb{2}H\sb{2} and HS species, because CEA does not carry the
thermodynamical parameters for them.

In the left panel of Figure \ref{fig:CEATEATEBS}, we see that for the
most species and temperatures CEA and TEA lines overlap (CEA result is
plotted in dashed and TEA in solid lines).  However, CH\sb{4} species
abundances above \math{T} \sim1700 K do not overlap.  TEBS colored
dots do not overplot either CEA or TEA curves, but follow them
closely.  This method is derived for only five major molecular species
and is based on a few simple analytic expressions.

In Figure \ref{fig:CEATEATEBS}, right panel, we again see that most
species overlap, except HCN and CH\sb{4}.  The HCN curves (for CEA and
TEA runs) differ for the full temperature range (see the inset
figure), while, as before, CH\sb{4} curve differs slightly only for
pressures above \sim0.1 bar and temperatures above \sim1700 K (see Figure 
\ref{fig:profiles} for the \math{T-P} profile used for this run). 

The differences seen in Figure \ref{fig:CEATEATEBS} come from the
different sources of thermodynamic data used for CEA and TEA 
(see Tables \ref{table:CEA-TEA-left} and \ref{table:CEA-TEA-right}). 
When the CEA thermodynamic data are used as input to TEA, all species final
abundances match, see Figure \ref{fig:CEATEA}. Section
\ref{sec:freeEner}, below, elaborates on this and investigate the difference
in free energy input values used for CEA and TEA.

\subsection{Comparison of free energy values in CEA and TEA}
\label{sec:freeEner}

The thermodynamic data used for CEA are in the form of polynomial
coefficients, and are listed in the {\tt termo.inp} file provided with
the CEA code. The format of this library is explained in Appendix A of
\citet{McBrideGordon:1996}. For each species, the file lists, among
other data, the reference sources of the thermodynamic data, the
values of the standard enthalpy of formation, \math{\Delta\sb{f}
  H\sb{298}\sp{0}}, at the reference temperature of 298.15 K and
pressure of 1 bar, and coefficients of specific heat, \math{C\sb{p}
  \sp{0}}, with integration constants for enthalpy, \math{H\sp{o}},
and entropy, \math{S\sp{o}}, for temperature intervals of 200 to 1000
K, 1000 to 6000 K, and 6000 to 20000 K.

The JANAF tables list the reference sources of their thermodynamic
parameters in \citet{ChaseEtal1986bookJANAFtables}. The data are also
available at {\tt http://kinetics.nist.gov/janaf/}. 

The difference in thermodynamic parameters between CEA and TEA is
noticeable even in their \math{\Delta\sb{f} H\sb{298}\sp{0}} values.
The source of standard enthalpies of formation in CEA for, e.g., HCN
and CH\sb{4} is \citet{gurvich1991criteria}, page 226 and 36,
respectively, and their respective values are 133.08 and -74.60
kJ/mol. The source of standard enthalpies of formation in the JANAF
tables is listed in \citet{ChaseEtal1986bookJANAFtables} on page 600
and 615, respectively, and their respective values are 135.14 and
-74.873 kJ/mol.

TEA uses JANAF tables to calculate the values of free energies for
each species following Equation \ref{eq:JANAFconv}.  To calculate the
values of free energies used in CEA, we started from Chapter 4 in
\citet{GordonMcBride:1994}. Our goal is to plug CEA free energies
into TEA and test whether TEA will produce the same final abundances
as CEA does.

As explained in Section 4.2, the thermodynamic functions specific
heat, enthalpy, and entropy as function of temperatures are given as:

\begin{eqnarray}
\frac {C_p^{o}}{R} = \sum\,a_i\,T^{q_{i}} \, ,
\label{Cp}
\end{eqnarray}

\begin{eqnarray}
\frac {H^{o}}{RT} = \frac{\int C_p^{o}\, dT}{RT} \, ,
\label{H}
\end{eqnarray}

\begin{eqnarray}
\frac {S^{o}}{R} = \int \frac{C_p^{o}}{RT}\, dT \, .
\label{S}
\end{eqnarray}

\noindent These functions are given in a form of seven polynomial
coefficients for specific heat, \math{C\sb{p}\sp{o}/R}, and two
integrations constants (\math{a\sb{8}} and \math{a\sb{9}}) for
enthalpy, \math{H\sp{o}/RT}, and entropy, \math{S\sp{o}/R}:

\begin{eqnarray}
\label{CpPoly}
\frac {C_p^{o}}{R} = a_1T^{-2} + a_2T^{-1} + a_3 + a_4T + a_5T^{2} +
a_6T^{3} + \\ \nonumber a_7T^{4} \, ,
\end{eqnarray}

\begin{eqnarray}
\label{HPoly}
\frac {H^{o}}{RT} = -\,\,a_1T^{-2} + a_2T^{-1}\,lnT + a_3 +
a_4\frac{T}{2} + a_5\frac{T^{2}}{3} + \\ \nonumber a_6\frac{T^{3}}{4}
+ a_7\frac{T^{4}}{5} + \frac{a_8}{T} \, ,
\end{eqnarray}

\begin{eqnarray}
\label{SPoly}
\frac {S^{o}}{R} = -\,\,a_1\frac{T^{-2}}{2} -\,a_2T^{-1} + a_3\,lnT +
a_4T + a_5\frac{T^{2}}{2} + \\ \nonumber a_6\frac{T^{3}}{3} +
a_7\frac{T^{4}}{4} + a_9 \, .
\end{eqnarray}

To derive free energies in the form that TEA uses them, we rewrite
Equation \ref{eq:JANAFconv} for one species as:

\begin{eqnarray}
\label{eq:JANAF}
\frac{g^0(T)}{RT} = 1/R\Big[\frac{G_{T}^0 - H_{298}^0}{T}\Big] +
\frac{\Delta_f H_{298}^0}{RT}\, ,
\end{eqnarray}

\noindent The first term on the right side can be expressed in the
following format \citep[][Page 3]{Chase1974janaf}:

\begin{eqnarray}
\label{eq:Chase1974}
\frac{G_{T}^0(T) - H_{298}^0}{T} = -\,\, S_T^{o} + \frac{(H_T^{o} -
  H_{298}^0)}{T} \, .
\end{eqnarray}

\noindent Thus, we rewrite Equation \ref{eq:JANAF} as:

\begin{eqnarray}
\label{eq:JANAF2}
\frac{g^0(T)}{RT} = 1/R\Big[\-\,\, S_T^{o} + \frac{(H_T^{o} -
    H_{298}^0)}{T}\Big] + \frac{\Delta_f H_{298}^0}{RT}\, ,
\end{eqnarray}

\begin{eqnarray}
\label{eq:JANAF2}
\frac{g^0(T)}{RT} = 1/R\Big[\-\,\, S_T^{o} + \frac{H_T^{o}}{T} -
  \frac{H_{298}^0}{T}\Big] + \frac{\Delta_f H_{298}^0}{RT}\, ,
\end{eqnarray}

\noindent To see Equations \ref{HPoly} and \ref{SPoly} inside Equation
\ref{eq:JANAF2}, we multiply and divide the first and second term on
the right with \math{R} and get:
 
\begin{eqnarray}
\label{eq:JANAF4}
\frac{g^0(T)}{RT} = \frac{S_T^{o}}{R} + \frac{H_T^{o}}{RT} -
\frac{H_{298}^0}{RT} + \frac{\Delta_f H_{298}^0}{RT}\, ,
\end{eqnarray}

In the CEA analysis paper, Section 4.1, \citet{GordonMcBride:1994}
state that they have arbitrary assumed \math{H\sp{o}(298.15) =
  \Delta\sb{f}H\sp{o}(298.15)}. Adopting this assumption leads to:

\begin{eqnarray}
\label{eq:JANAF4}
\frac{g^0(T)}{RT} = \frac{S_T^{o}}{R} + \frac{H_T^{o}}{RT} -
\frac{\Delta_f H_{298}^0}{RT} + \frac{\Delta_f H_{298}^0}{RT}\, .
\end{eqnarray}

\noindent The last two terms cancel leading to a simple expression for
free energies:

\begin{eqnarray}
\label{eq:JANAF4}
\frac{g^0(T)}{RT} = \frac{S_T^{o}}{R}  + \frac{H_T^{o}}{RT}\, ,
\end{eqnarray}

\noindent The first term on the right side is Equation \ref{SPoly}, while
the second term is Equation \ref{HPoly}; expressions with polynomial
coefficients that are given in the CEA {\tt thermo.inp} file.

Following the last conclusion, we calculated the free energies for each
species of interest and used them as input to TEA.  Figure
\ref{fig:CEATEA} shows the comparison between CEA and TEA using CEA
free energies. We see that all species overlap. Tables
\ref{table:CEA-TEA-left} and \ref{table:CEA-TEA-right} give the exact
values of free energies used and the final abundances
for several (\math{T, P}) points that showed the largest differences
between CEA and TEA runs in Figure \ref{fig:CEATEATEBS}. It also lists
the free energies calculated using JANAF tables and the final
abundances produced by TEA using JANAF thermodynamic data.

As seen in Figure \ref{fig:CEATEA}, although CEA uses Newton-Raphson
and TEA the Lagrangian method of steepest descent, both approaches,
using the same inputs (free energies), find the same final
abundances. Table \ref{table:CEA-TEA-left}, (groups {\em CEA final
  abundances} and {\em TEA final abundances using CEA free energies}),
shows values identical for most species between the two tests.  A few
cases show that abundance ratios are inconsistent at the 10\sp{-5}
level. Table \ref{table:CEA-TEA-right} displays the same trend. The
differences in the fifth decimal place may indicate that, somewhere in
CEA, a calculation is carried out in 32-bit precision, possibly due to
a literal single-precision number in the source code.  Python floating
literals are in 64-bit precision by default.

\section{Reproducible Research License}
\label{sec:RR}

Reproducing a lengthy computation, such as that implemented in TEA,
can be prohibitively time consuming \citep{stodden2009legal}.  We have released TEA under an
open-source license, but this is not enough, as even the most
stringent of those licenses (e.g., the GNU General Public License)
does not require disclosure of modifications if the researcher does
not distribute the code.  So that the process of science can proceed
efficiently, there are several terms in our license to ensure
reproducibility of all TEA results, including those from derivative
codes. A key term requires that any reviewed scientific publication using TEA
or a derived code must publish that code, the code output used in the
paper (such as data in tables and figures, and data summarized in the
text), and all the information used to initialize the code to produce
those outputs in a reproducible research compendium (RRC).  The RRC must
be published with the paper, preferably as an electronic supplement, or
else in a permanent, free-of-charge, public internet archive, such as
github.com.  A permanent link to the archive must be published in the
paper, and the archive must never be closed, altered, or charged for.
Details and examples of how to do this appear in the license and
documents accompanying the code, along with additional discussion.  The
RRC for this paper, including the TEA package and documentation, is
included as an electronic supplement, and is also available via
{\tt https://github.com/dzesmin/ RRC-BlecicEtal-2015a-ApJS-TEA/}.

\section{Conclusions}
\label{sec:conc}

We have developed an open-source Thermochemical Equilibrium Abundances
code for gaseous molecular species. Given elemental abundances and one
or more temperature-pressure pairs, TEA produces final mixing
fractions using the Gibbs-free-energy minimization method with an
iterative Lagrangian optimization scheme.

We applied the TEA calculations to several hot-Jupiter \math{T-P}
models, with expected results. The code is tested against the
original method developed by \citet{WhiteJohnsonDantzig1958JGibbs},
the analytic method developed by \citet{BurrowsSharp1999apjchemeq},
and the Newton-Raphson method implemented in the free Chemical
Equilibrium with Applications code. Using the free energies listed in 
\citet{WhiteJohnsonDantzig1958JGibbs}, their example, and derived free
energies based on the thermodynamic data provided in CEA's {\tt
  thermo.inp} file, TEA produces the same final
abundances, but with higher precision.

Currently, TEA is specialized for gaseous species, with the
implementation of condensates left for future work. In opacity
calculations at low temperatures (below 1000 K), the inclusion of
condensates is necessary as it reduces the gas phase
contribution to opacity \citep[e.g., ][]{SharpHuebner90, Lodders02,
  BurrowsSharp1999apjchemeq}.

The thermochemical equilibrium abundances obtained with TEA can be
used in all static atmospheres, atmospheres with vertical transport
and temperatures above 1200 K (except when ions are present), and as a
starting point in models of gaseous chemical kinetics and abundance
retrievals run on spectroscopic data. TEA is currently used to
initialize the atmospheric retrial calculations in the open-source
BART project (available at {\tt https://github.com/joeharr4/BART}).

TEA is written in a modular way using the Python programming
language. It is documented (the Start Guide, the User Manual, 
the Code Document, and this theory paper are provided with the code), 
actively maintained, and available to the community via the open-source
development sites {\tt https://github.com/dzesmin/TEA} and
{\tt https://github.com/dzesmin/TEA-Examples}.

\acknowledgments This project was completed with the support of the
NASA Earth and Space Science Fellowship Program, grant NNX12AL83H,
held by Jasmina Blecic, PI Joseph Harrington, and through the Science
Mission Directorate's Planetary Atmospheres Program, grant
NNX12AI69G. We would like to thank Julianne Moses for useful
discussions, and Kevin B. Stevenson and Michael R. Line for
temperature and pressure profiles. We also thank contributors to
SciPy, NumPy, Matplotlib, and the Python Programming Language; the
open-source development website GitHub.com; and other contributors to
the free and open-source community. \\

\bibliography{TEA}

\end{document}